\documentclass[authoryear,3p]{elsarticle}
\usepackage{natbib, epsfig, amssymb, multirow}

\begin{document}

 \begin{frontmatter}
\title{The Vertical Structure of the F Ring of Saturn from Ring-Plane Crossings}
\author{Britt R. Scharringhausen\corref{cor1}}
\ead{scharr@beloit.edu}
\address{Department of Physics and Astronomy, Beloit College, Beloit, WI, 53511}
\cortext[cor1]{Corresponding author: \\
Britt Scharringhausen\\
Beloit College\\
Beloit, WI 53511\\
Email: scharr@beloit.edu\\
Phone: (608) 363-2648 \\
Fax: (608) 363-2052
}

\author{Philip D. Nicholson}
\address{Department of Astronomy, Cornell University, Ithaca, NY, 14853}
\ead{nicholso@astro.cornell.edu}
\begin{abstract}

  We present a photometric model of the rings of Saturn which includes
  the main rings and an F~ring, inclined to the main rings, with a
  Gaussian vertical profile of optical depth.  This model reproduces
  the asymmetry in brightness between the east and west ansae of the
  rings of Saturn that was observed by the Hubble Space Telescope
  (HST) within a few hours after the Earth ring-plane crossing (RPX)
  of 10 August 1995. The model shows that during this observation the
  inclined F ring unevenly blocked the east and west ansae of the main
  rings. The brightness asymmetry produced by the model is highly
  sensitive to the vertical thickness and radial optical depth of the
  F~ring.  The F-ring model that best matches the observations has a
  vertical full width at half maximum of $13 \pm7$ km and an
  equivalent depth of $10 \pm 4$ km.  The model also reproduces the
  shape of the HST profiles of ring brightness vs. distance from
  Saturn, both before and after the time of ring-plane crossing.
  Smaller asymmetries observed before the RPX, when the Earth was on
  the dark side of the rings, cannot be explained by blocking of the
  main rings by the F~ring or vice versa and are probably instead due
  to the intrinsic longitudinal variation exhibited by the F~ring.

\end{abstract}

\end{frontmatter}

\section{Introduction}

With a semimajor axis of 140,200~km \citep{B02}, the F~ring of Saturn
lies $\sim 3500$ km outside the A ring.  The narrow F~ring is largely
diffuse and dusty, composed of particles $\sim 0.5 \mu$m in size, with
a highly variable optical depth that is generally $< 1$.  It may have
a narrow core of greater optical depth, composed of macroscopic
particles \citep{S92}. Because the F~ring lies just outside the main
rings and is only weakly backscattering, it is usually not observable
from Earth.  An exception occurs during a ring-plane crossing (RPX),
when the small ring-opening angle drastically reduces the brightness 
of the main rings to levels comparable to, or less than, the
brightness of the F ring.

The Sun crosses Saturn's ring plane during the equinoxes of Saturn,
approximately every 15 years. As Saturn's ring plane sweeps through
the inner Solar System, Earth's orbital motion carries it from one
side of Saturn's ring plane to the other.  During each
ring-plane-crossing season, which takes less than 12 months, there is
one solar RPX.  In about half of RPX seasons, the Earth crosses the ring
plane three times, with one Earth ring-plane crossing occurring near opposition; the last
such event occurred on 10 August 1995.  In the other half there is
only one Earth ring-plane crossing which occurs near solar conjunction
and is thus generally not observable from Earth (as was the case in
the RPX of 2009).

An Earth RPX gives us an excellent opportunity to probe the F~ring's
properties.  During an RPX, the observed brightness of the main rings
is greatly reduced but because the F ring is optically thin, its brightness is unaffected.  When the Earth is
on the dark side of the main rings near RPX, the F ring dominates the
overall brightness of the rings.  Even when Earth is on the lit side
of the rings, the brightness of the F~ring is comparable to the brightness of the main rings when the ring-opening angles to Earth and the Sun are small.  Due to its vertical thickness, the F~ring also can partially obscure the main rings and so can block some of their reflected sunlight.   This means that the overall brightness of the rings during an Earth ring-plane crossing cannot be understood without understanding the role of the F~ring. 

Early observers were stymied when attempting to measure the thickness of Saturn's rings during previous Earth ring-plane crossings, finding thicknesses much greater than measurements of the main ring thickness derived from theoretical estimates and Voyager observations.  An asymmetry in brightness between the east and west ansae near the Earth ring-plane crossing of 10 August 1995  also could not be explained using the known properties of the main rings. In this paper we will show that the F~ring provides the solution to both of these puzzles.

\subsection{Early estimates of ring thickness}
Historically, an Earth RPX has been seen as an opportunity to measure
the vertical thickness of the rings as they are seen edge-on.  The
vertical thickness of the rings is too small be resolved,
even by spacecraft \citep{G78,E84,Z85,S03}.  The projected height
of the rings on the sky is small enough that from Earth the rings are
vertically unresolved for weeks before and after the RPX. Prior to
Voyager, the residual ring brightness at the moment of the RPX was
generally assumed to be due to light reflected from the outer edge of
the A ring and possibly the edge of the B ring seen through the Cassini
Division \citep{B81}. 

The brightness of the rings, expressed as an equivalent photometric thickness with units of length, was traditionally interpreted using radiative transfer theory to estimate the physical thickness of the rings.
Based on various observations of the 12--13 June 1966 RPX, the
photometric thickness of the rings was found to be $0.57$~km
\citep{B72}, $0.8\stackrel{+2.3}{\scriptstyle{-0.8}}$~km \citep{L79},
$1.3\pm0.3$~km \citep{F78}, and $2.4\pm1.3$~km \citep{D79}.  These
measurements were taken from photographic plates and the ring's edge-on
reflectance was assumed to be similar to the reflectance of Saturn.
By modeling the rings as a
plane-parallel scattering layer of finite thickness, \citet{Si82} found a thickness of
$1.1\stackrel{+0.9}{\scriptstyle{-0.5}}$~km using observations of the RPX of 12 March 1980.

These photometric ring thicknesses are significantly larger than measurements of the main ring thickness obtained by interpreting various types of spacecraft observations.

\subsection{Voyager observations of ring thickness}
An occultation of the star $\delta$ Scorpii by the rings was observed
using the Voyager 2 Photopolarimeter Subsystem (PPS).  By measuring
the rate of change of the transmitted flux of the occulted star as it
crossed several sharp transitions from opaque to transparent regions of the
main rings, an upper limit of 200~m was placed on the thickness of the
main rings \citep{L82}.
Measurements of the
forward-scattering of microwaves transmitted through the rings from
Voyager 1 indicated a thickness of 10~m in the C ring and 20--50~m in
the A ring \citep{Z84} although the results are somewhat model-dependent.

Based on the damping distance of density and bending waves in the
rings, \citet{E84} argued for a main-ring thickness of
30--35~m. \citet{G78} modeled the ring particles as inelastically
colliding spheres and concluded that, if the rings have come to an
equilibrium between viscous dissipation and stirring by collisions,
their vertical thickness is likely to be less than 10~m.  If less
elastic collisions are assumed, the model implies that the rings will
collapse to a monolayer.

\citet{S03} combined a sophisticated dynamical model of the rings with
a Monte Carlo photometric model. The B ring's
increased brightness at opposition (the opposition effect) and
increasing brightness at larger opening angles (the tilt effect)
were best reproduced by a model with a distribution of particle sizes
from 10~cm to 5~m, although this lower limit is slightly too high to
be consistent with the particle size distribution derived from the
Voyager radio occultation experiment \citep{Z85}. In this model, the
larger particles settled into a near-monolayer, though the smaller
particles had a vertical distribution of a few tens of meters.

While occultation measurements and dynamical models of the rings vary
in their estimates of the thickness of the rings, none of these models
provides a ring thick enough to account for the residual brightness of
the rings during an Earth RPX, although \citet{L84} found that the
Mimas 5:3 bending wave in the A~ring has a vertical amplitude of
~500~m, which may account for part of the observed edge-on brightness.

\subsection{The 1995 ring-plane crossings}

Assuming a geometric albedo of 1 for the rings, \citet{B97} estimated
an equivalent thickness of $1.4\pm0.1$~km from Hubble Space Telescope (HST) images taken during the Earth
ring-plane crossing of 22 May 1995, while \citet{N96} found a
vertically-integrated reflectance of $1.22\pm0.17$ km for the east ansa and $1.53\pm0.09$ km for the west ansa.  (These figures do not 
suffer from the calibration problem to be discussed in Sec.~\ref{sec:vif}.)
Both of these thicknesses are similar to earlier ground-based photometric thicknesses.

Profiles of brightness as a function of radius from the set of HST images spanning the
August 1995 Earth RPX showed two surprising features.  In images of the
dark side of the rings, when one would expect very little light from
the main rings, the profile of ring brightness as a function of
distance from the center of Saturn (see
Fig.~\ref{plot:hst_profiles_e_w}) is fairly uniform out to the
semimajor axis of the F~ring, where it drops rapidly, suggesting
that the primary contribution to the ring brightness in this geometry
comes from the F~ring (rather than the Mimas bending wave at $\sim$133,000 km).  A more surprising discovery was that the
radially-averaged brightness of the rings was greater on the west ansa
than the east ansa, with this asymmetry reversing $\sim$3 hours after
the ring-plane crossing. \citep{N96}

In HST observations of the dark side of the rings at an Earth ring-opening angle
of $B_e=2.67^\circ$ obtained during the November 1995 solar RPX, the F~ring is prominent at the edge of the main rings, but disappears
$\sim35^\circ$ to the rear of the east ansa, suggesting that the
F~ring is inclined and that it was being shadowed by the main rings
\citep{B96, N96}.

We exploit this asymmetry to find the equivalent
depth and vertical extent of the F~ring.

\section{HST observations of  the 10 Aug 1995 ring-plane crossing}

HST observed the Earth ring-plane crossing on 10 August 1995. The
0.89~$\mu$m methane filter (FQCH4N) was used to reduce scattered light
from Saturn without suppressing the brightness of the rings.  HST's
orbit around the Earth has a 96-minute period so that ``visits,'' or
observing opportunities, were $\sim$20--40 minutes in duration.  We group the images
into seven visits which we label with the round times 14:00, 15:30, 18:30, 20:00,
22:00, 23:30, and 25:00.  All times are given in UT. The last time
is actually 1:00 UT on 11 August 1995.  From 14--20 UT, the Earth 
was on the dark side of the main rings, and from 22--25 UT the Earth was on
the sunlit side.

In each visit, images were taken with the Planetary Camera (PC) and
the Wide-Field Camera (WF3), which were the first and third chips,
respectively, of the Wide-Field Planetary Camera 2
(WFPC2). Table~\ref{table:ring_opening} gives the approximate times of
each of these HST visits and the ring-opening angles at these
times. The pixels in the WF3 images have a size of 634.2 km at Saturn, and the
images include both ring ansae.  PC images, at 290.0 km/pixel, have a
smaller field of view, and contain only one ansa (the east ansa until
23:00, and the west ansa thereafter). For further details of the
observation sequence, image numbering, etc., the reader is referred to
\citet{M01}.

\subsection{HST profiles}

\begin{table}
\caption{Earth and Sun ring-opening angles ($B_e$ and $B_s$,
  respectively) for HST observations, interpolated from an ephemeris
  provided by the Planetary Data System's Rings Node.}
\label{table:ring_opening}
\begin{center}
\begin{tabular}{cr@{:}lr@{.}lr@{.}l}
\hline
\multicolumn{3}{c}{UT Date and Time} & \multicolumn{2}{c}{$B_e $($^\circ$)} 
& \multicolumn{2}{c}{$B_s$($^\circ$)} \\ \hline
1995 Aug. 10 & 14&00 & $-$0&0078 & 1&498 \\
1995 Aug. 10 & 15&30 & $-$0&0061 & 1&497 \\
1995 Aug. 10 & 18&30 & $-$0&0026 & 1&496 \\
1995 Aug. 10 & 20&00 & $-$0&0008 & 1&495 \\
1995 Aug. 10 & 22&00 &    0&0015 & 1&493 \\
1995 Aug. 10 & 23&30 &    0&0031 & 1&492 \\
1995 Aug. 11 &  1&00 &    0&0050 & 1&492 \\
1995 Nov. 21 & 14&00 &  2&670 & $-$0&0290\\
\hline
\end{tabular}
\end{center}
\end{table}

During the HST observations, the projected ``height'' of the main ring
($2 R_A \sin{B_e}$, where $R_A$ is the outer radius of the A ring) was
$\lesssim 40$ km, much less than the image resolution of $0.1$
arcsec$~ \approx ~$650~km, so the rings are unresolved in the vertical
dimension and remain so for $|B_e|<0.14^\circ$, approximately 4 days
before and after the August 1995 RPX.  The ring-opening angles for
each visit are given in Table~\ref{table:ring_opening}.  Profiles were
produced by rotating HST images to make the rings horizontal and
summing columns of pixels perpendicular to the rings to find the
vertically-integrated I/F as a function of $r$, the horizontal
distance from the center of Saturn:
\begin{equation}
\rm{VIF(r)}=\int \rm{I/F} dz=\sum_z \langle {\rm I/F} \rangle_{pix} \sigma_z.
\label{eq:vif}
\end{equation}
Here, $\sigma_z$ is the projected vertical dimension of a pixel in the plane of the
sky at the distance of the observed rings. $\langle {\rm I/F}
\rangle_{pix}$ is the mean reflectance of the rings for that pixel, computed
from the brightness of the pixel obtained using the standard HST
calibration pipeline, the solar spectrum, and the relative distances of the Earth, Sun, and
Saturn.  We also subtracted a sky level obtained by averaging the rows
of pixels above and below the rings in order to remove significant
levels of scattered light from Saturn.

All of the resulting HST profiles are contaminated by light from satellites. Because their orbital motion carries the satellites a significant distance over the course of a single HST visit, it is often possible to remove much of their contribution by median-filtering the profiles for the visit.  The profiles extracted from WF3 and PC images are sampled at intervals of 640~km and 290~km/pixel, respectively, and we median filtered all the available profiles of each ansa from each HST visit to produce composite profiles with a resolution of 650~km.  If there were only two images of a given ansa in that visit, we took the {\it
  minimum} of the two profiles at each pixel as the value for the composite profile.
This removes many of the satellites, but fails to remove all traces of
bright satellites, satellites near the ansa (where their motion in $r$
is slower), or instances where two or more satellites were close to
one another in $r$. Dips around $r=$105,000~km in the 25:00 UT profile
are due to flux from Mimas, which has a noticeable inclination, and
thus contaminates the sky subtraction.

\begin{figure} 
\begin{center}
\includegraphics[width=3.8in]{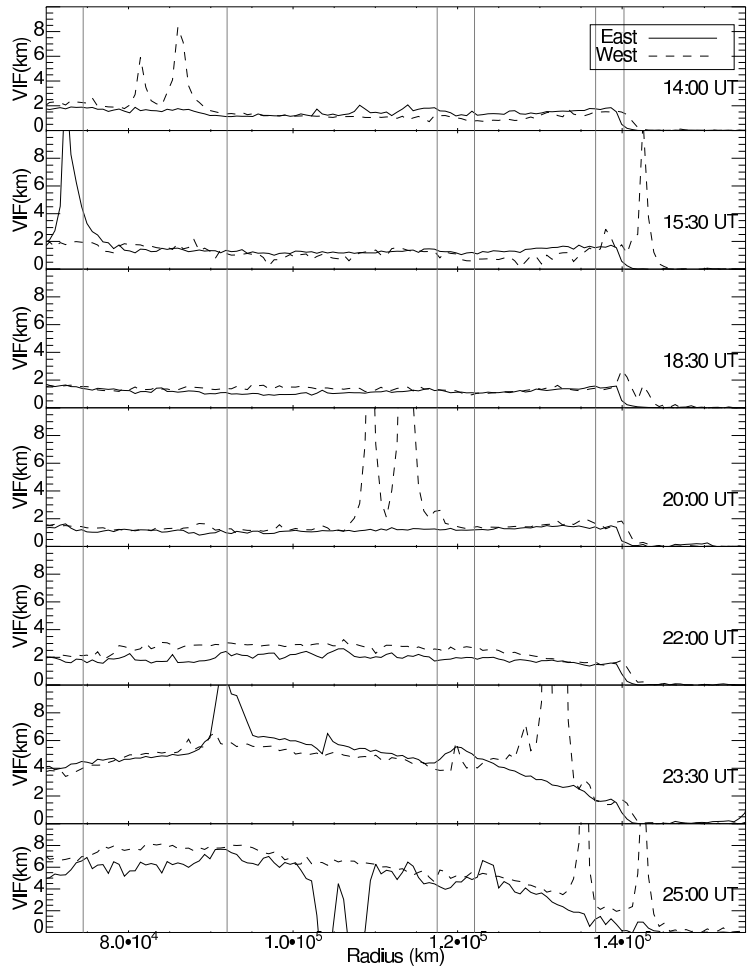}
\end{center}
\caption[Profiles of VIF($r$) extracted from HST images.]{Profiles of
  VIF($r$) extracted from images of the east and west ansae of the
  rings from HST images taken on 10--11 August 1995. All the profiles
  are plotted on the same vertical scale. The light from satellites
  was removed through median or minimum filtering of all the profiles
  from each ansa in each HST visit, but the remnants of some
  satellites remain as valleys (where satellite light contaminated the
  sky subtraction) or peaks. The vertical lines indicate the
  inner border of the C~ring, the boundary between the C and B~ring,
  the inner and outer borders of the Cassini Division, the outer edge
  of the A ring, and the location of the F~ring.}
\label{plot:hst_profiles_e_w}
\end{figure}

The resulting profiles, published for the first time here, are shown
in Fig. \ref{plot:hst_profiles_e_w}. When viewed on the dark side
(14:00--20:00 UT) and nearly edge-on (22:00 UT), the rings are uniform
in brightness out to a radius beyond the outer edge of the A ring
(136,800~km) but drop off steeply at the location of the F~ring
(140,200~km).  As originally suggested by \citet{N96}, this implies
that the edge-on brightness is dominated, not by light reflected from
the vertical edge of the main rings, as had generally been assumed,
nor from the Mimas 5:3 bending wave, but from the F~ring. After the
ring-plane crossing, the brightness of the main rings increases
dramatically.

These profiles were produced to illustrate the variation of VIF with
radius.  Due to inconsistencies in the calibration between WF3 and PC
chips (see Sec.~\ref{sec:vif}) and the incomplete removal of flux from
satellites, the overall brightness is not well-determined and the
asymmetry between the ansae is not clearly shown in these profiles.

\subsection{HST $\langle$VIF$\rangle$ vs. time}

\begin{figure} 
\begin{center}
\includegraphics[width=5in]{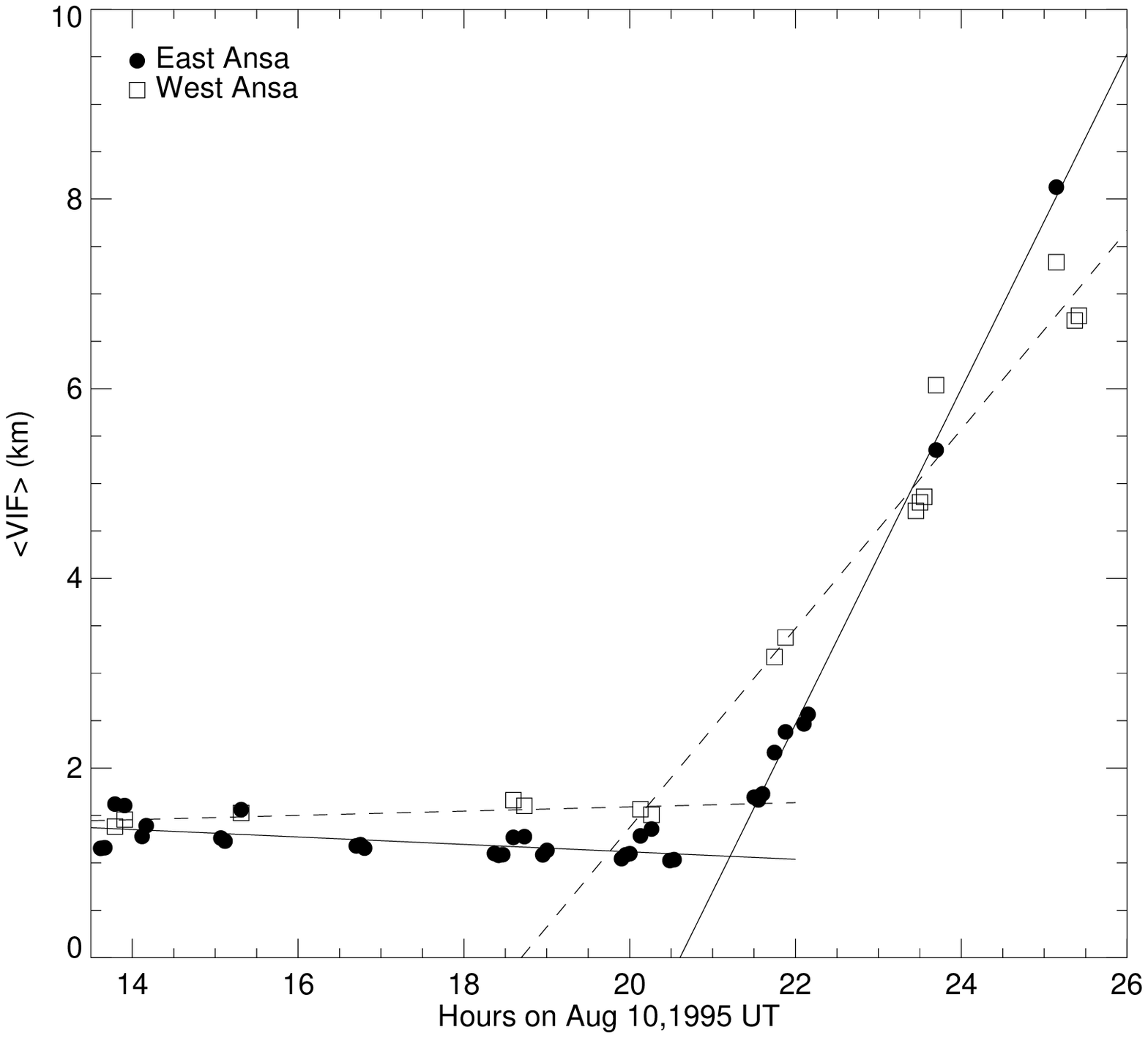}
\end{center}
\caption{The evolution of the brightness of the rings during the Earth
  ring-plane crossing of 10 August 1995 as observed by HST. This plot
  shows data from both WF3 and PC chips, which have been recalibrated (cf. Fig. 2 of \citet{N96}). The lines are linear fits
  to the data from the east and west ansae, before and after the
  ring-plane crossing.  Their intersection gives the
  ring-plane-crossing time, which we find to be 20:13$\pm$22 min for the west
  ansa and 21:13$\pm$3 min for the east ansa. }
\label{plot:hst_correct}
\end{figure}

We have reproduced the ring brightnesses measured by \citet{N96} by averaging the VIF of the rings over a range of
$r=80,000$--120,000~km, after removing light from the satellites.
This $\langle$VIF$\rangle$ is plotted vs. time in
Fig.~\ref{plot:hst_correct}.  The intersection of linear fits to the
$\langle$VIF$\rangle$ of each ansa before and after the ring-plane
crossing was used to compute the ring-plane-crossing time for each
ansa.  Uncertainties were calculated from the statistical
uncertainties in the fit parameters.

\label{sec:vif}
These data were taken with the narrow-band methane filter at 890~nm,
which was part of a unique ``quad filter'' in WFPC2 that was used only
infrequently and was rather poorly calibrated at the time of the
observations.  The figure presented here differs slightly
from Fig. 2 of \citet{N96} because a photometric calibration factor,
which differed slightly between the WF3 and PC images, was not applied
correctly to the data plotted in the original work, where all data
were mistakenly scaled by the same factor.

The ring-plane-crossing times originally computed from these incorrectly
calibrated data by ~\citet{N96} were 20:20 UT $\pm$ 8 min for the west ansa and 21:09
UT $\pm$ 2 min for the east ansa.  With the corrected photometric
calibration, we find slightly revised ring-plane-crossing times of
20:13$\pm$22 min for the west ansa and 21:13$\pm$3 min for the east
ansa.

\citet{N96} also averaged over all pre-crossing 
data points and found an average dark side $\langle$VIF$\rangle$ of
1.53$\pm$0.09~km on the west ansa and 1.22$\pm$0.17~km on the east
ansa.  It appears that the data used in this calculation were
calibrated correctly.

The mean values of the $\langle$VIF$\rangle$ of each ansa for each HST
visit are given in Table~\ref{table:HST_VIF}.  When there is more than
one image of an ansa per visit, uncertainties are computed from the
scatter in data points. (\citet{N96} do not quote any uncertainties
for individual measurements.)

\begin{table}
\caption{Average HST visit $\langle$VIF$\rangle~$and asymmetries for
  each visit 10--11 August 1995.  Uncertainties were computed from the
  scatter in data points for each visit, unless only one image was taken.}
\label{table:HST_VIF}
\begin{center}
\begin{tabular}{clllr}
\hline
     & UT Time & $\langle$VIF$\rangle_E$ (km)  & $\langle$VIF$\rangle_W$ (km) & $\Delta\langle$VIF$\rangle$ (km) \\ 
\hline
     &   14:00 & 1.37 $\pm$ 0.19 & 1.42 $\pm$ 0.04 & -0.05 $\pm$ 0.19 \\
     &   15:30 & 1.35 $\pm$ 0.15 & 1.53            & -0.18 $\pm$ 0.21 \\
Dark &   16:30 & 1.18 $\pm$ 0.02 & \multicolumn{1}{c}{---} &\multicolumn{1}{c}{---}  \\ 
     &   18:30 & 1.15 $\pm$ 0.08 & 1.63 $\pm$ 0.03 & -0.48 $\pm$ 0.09 \\
     &   20:00 & 1.13 $\pm$ 0.12 & 1.54 $\pm$ 0.03 & -0.41 $\pm$ 0.12 \\
\hline
     &   22:00 & 2.09 $\pm$ 0.36 & 3.27 $\pm$ 0.10 & -1.18 $\pm$ 0.37 \\
Lit  &   23:30 & 5.35            & 5.10 $\pm$ 0.54 &  0.25 $\pm$ 0.75 \\
     &   25:00 & 8.13 	         & 6.94 $\pm$ 0.28 &  1.19 $\pm$ 0.40 \\ 
\hline
\end{tabular}
\end{center}
\end{table}

Table~\ref{table:HST_VIF} also gives the asymmetry in brightness between the east and west ansae, defined as
\begin{equation}
\Delta\langle\rm{VIF}\rangle=
	\langle\rm{VIF}\rangle_E-\langle{\rm VIF}\rangle_W.
\end{equation}
In the 16:30~UT visit, only the east
ansa was imaged, so no asymmetry is given. When the uncertainty can
only be computed for one ansa, we estimate the error in the asymmetry
in $\langle$VIF$\rangle$ by multiplying that error by $\sqrt 2$.

In Fig.~\ref{plot:hst_correct}, in the observations of the dark side
of the rings, $\langle$VIF$\rangle_W$, was slightly greater than
$\langle$VIF$\rangle_E$, with the asymmetry increasing as the RPX
approached. It is not clear if the difference in brightness between
the ansae is statistically significant before 18:30~UT.

For the first observations after the ring-plane crossing, at 22:00
UT, the west ansa is brighter than the east ansa by 56\%.  The sense of
the asymmetry reverses around 24:00 UT, and in the last set of data,
the east ansa is brighter than the west ansa by 16\%.

\begin{figure} 
\begin{center}
\includegraphics[width=5in]{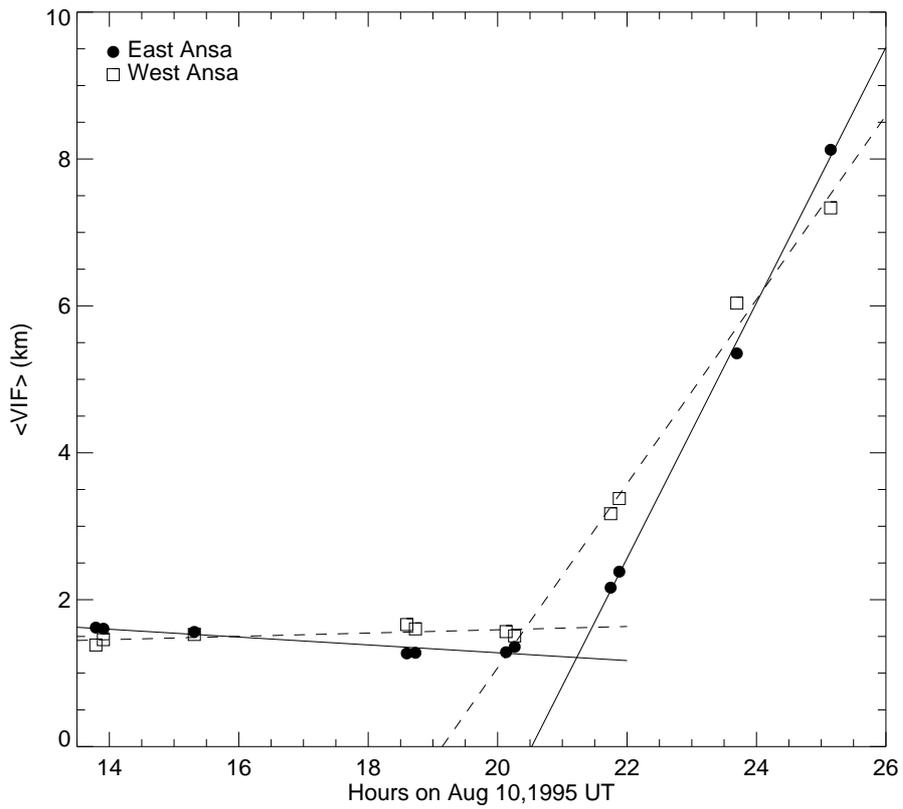}
\end{center}
\caption[HST $\langle$VIF$\rangle$ plotted vs. time, including WF3
images only.]{The evolution of the mean brightness of the rings during the
  Earth ring-plane crossing of 10 August 1995 as observed by HST. This
  plot shows data from only the WF3 chip
  (cf. Fig.~\ref{plot:hst_correct} which contains data from both
  chips).  The lines are linear fits to the data.  The intersection of
  linear fits for each ansa before and after the ring-plane crossing
  give an estimate of the ring-plane-crossing time. Using WF3 data
  only, we find the RPX times to be UTC 20:25$\pm$10~min for the west ansa and  UTC 21:13$\pm$3 min for the east ansa.}
\label{plot:hst_wf3}
\end{figure}

In order to avoid bright satellites, the PC imaged only the east ansa before 23:00 UT
and only the west ansa after 23:00 UT, and the relative calibration of the WF3 and PC images remains a matter for
concern.   For these reasons, we will be
comparing our model results only to results from WF3, because these
images contain both ansae, and thus the relative calibration between
east and west ansae is more certain.  Fig.~\ref{plot:hst_wf3} is a
plot of $\langle$VIF$\rangle$ vs. time including only WF3 images.
Computing the RPX times as before, we find the east ansa crossing
time to be 21:13$\pm$3 min and the west ansa crossing time to be
20:25$\pm$10 min. Table~\ref{table:HST_VIF_WF3} gives the asymmetries
calculated from the WF3 images alone for each HST visit.  Notice that
these show that the {\it east} ansa is slightly brighter before 16:00
UT, in contrast to Fig.~\ref{plot:hst_correct} where the west ansa is
as bright or brighter than the east ansa at all times before the RPX.

\begin{table}
\caption{Radially averaged VIF for WF3 images only on 10 August 1995.}
\label{table:HST_VIF_WF3}
\begin{center}
\begin{tabular}{rrrr}
\hline

UT Time & $\langle$VIF$\rangle_E$ & $\langle$VIF$\rangle_W$ & 
$\Delta\langle$VIF$\rangle$ \\
 & (km) & (km) & (km) \\
\hline
 14:00 &  1.61  &     1.41  &    0.19 \\
 15:30 &  1.56  &     1.53  &    0.03 \\
 18:30 &  1.28  &     1.63  &   -0.35 \\
 20:00 &  1.32  &     1.54  &   -0.22 \\
 22:00 &  2.27  &     3.28  &   -1.01 \\
 23:30 &  5.35  &     6.04  &   -0.69 \\
 25:00 &  8.13  &     7.33  &    0.80 \\
\hline

\end{tabular}
\end{center}
\end{table}

Our model seeks to reproduce both the shapes of the HST profiles in Fig.~\ref{plot:hst_profiles_e_w} as well as the asymmetries in Fig.~\ref{plot:hst_wf3}.

\section{The F~ring}

\subsection{Spacecraft imaging observations of the F~ring}

The F~ring was discovered
in Pioneer 11 images, where the F~ring's clumpy nature
was immediately apparent \citep{G80}.  Voyager 1 and 2 revealed up to
four strands in the core of the F~ring that were braided and kinked at
some longitudes \citep{S81, Sm82}.  Clumps were observed to maintain
their integrity for up to 15 orbits around Saturn \citep{S04}. 

In Voyager Wide-Angle Camera images, the F~ring typically varies in
brightness by a factor of four with longitude in large part due to phase angle variations, 
but with comparable variations due to clumping \citep{S92}.

The Cassini Imaging Science Subsystem (ISS) has obtained complete
azimuthal coverage of the F~ring, revealing that in the current epoch
the F~ring has a a radial width of $\sim$700 km.  The F ring has a
core with a radial width of $\sim$20 km. \citep{M04}.

Longitudinally-complete F-ring
observations by ISS at one orbital position revealed that
the strands and core of the F~ring exhibit complex, temporally
evolving structure due to interactions with Prometheus \citep{K90, M05, B10} and
smaller clumps and moonlets in the F-ring region \citep{Ba02}. \citet{C05}
found that the strands of the F~ring, as observed November 2004--May
2005, constitute a single spiral arm.  This kinematic spiral may have
been created by the collision of a kilometer-sized moonlet with the
F~ring \citep{M08}.

\citet{M08} suggest that there may be larger particles
distributed irregularly in the F ring, because moonlets passing
through the F ring produce features through impacts, but do not do so
with every pass.  While they may be in large part responsible for the
F-ring's variable structure, larger objects within the F ring and other
moonlets near the F ring do not scatter a large amount of light
in the near infrared, and compared to the more extended dusty core and
dust envelope, do not contributed significantly to the equivalent
depth, as discussed below.

\subsection{Occultation measurements of the  F~ring}

\label{sec:occultations}

\begin{table}
\caption{Equivalent depths (D) of the F~ring measured from stellar
  occultations in the visible, infrared, and ultraviolet.}
\label{table:equivalent_depths}

\begin{minipage}{\textwidth}
\begin{tabular}{ l l r r @{$\pm$} l r l l}
Star & Year & B ($^{\circ}$) &\multicolumn{2}{c}{D (km)} & Wavelength & Observatory  & References \\
\hline
$\delta$ Sco                                        & 1981                           &                           28.3 & 4.33 & 0.13 & 0.264 $\mu$m & Voyager 2 &  {\emph a} \\ 
\hline                                                     
\multirow{4}{*}{28 Sgr}                        &  \multirow{4}{*}{1989} & \multirow{4}{*}{25.4} & 3.79 & 0.08 & 3.1 $\mu$m    & IRTF          &  {\emph b}\\
                                                            &                                    &                                   & 3.0   & 0.1   & 3.9 $\mu$m    & Palomar    &  {\emph b} \\
                                                            &                                    &                                   & 2.8   & 0.1   & 2.3 $\mu$m    & MacDonald (ingress)    & {\emph c} \\
                                                            &                                    &                                   & 3.6   & 0.1   & 2.3 $\mu$m    & MacDonald (egress)    & {\emph c} \\
\hline
GSC5249         & \multirow{2}{*}{1995} & \multirow{2}{*}{-2.67}& 7.41 & 0.15 & 270--740 nm & HST & {\emph c} \\
~~~-01240                                                            &                                    &                                   & 5.76   & 0.06   & 2.3 $\mu$m    & IRTF    & {\emph c} \\
\hline
Various                                                & 2005--8 & Various                      & 9.7 &  & 110--190 nm  & Cassini & {\emph d} \\
Various                                                & 2005--9 & Various                      & 9.99 &  2.65  & 2.92 $\mu$m  & Cassini & {\emph e} \\

\hline

\end{tabular}
\footnotetext[1]{\citet{L84, S92, N90,A12}} 
\footnotetext[2]{\citet{N00}} 
\footnotetext[3]{\citet{B02}} 
\footnotetext[4]{\citet{A12}} 
\footnotetext[5]{\citet{H11, F11}} 

\end{minipage}

\end{table}

The emerging picture of the F ring as revealed by occultation studies
is that it consists of a large, diffuse envelope of dusty material with an optically thicker dusty
core, sometimes accompanied by dusty strands.

An occultation of the star $\delta$ Scorpii by the rings, observed at 0.264~$\mu$m
by the Voyager 2 PPS, yielded a profile
of vertical optical depth, $\tau_v(a)$, measured along a line of sight perpendicular to the ring plane, as a function of $a$, the radial distance from the
center of Saturn.  In the F-ring, an often-stranded core $\sim3$~km wide had an optical depth $\sim
$0.5 and $\sim$1 in some narrow ($<0.5~$km) regions
\citep{L82}. In this profile, the envelope extends at least 50~km
radially, and the total equivalent depth,
\begin{equation}
{\rm D}=\int \tau_v(a)da,
\label{eq:equivalent_depth}
\end{equation}
was measured to be 4.33$\pm$0.13~km \citep{S92}.  The ring-opening
angle to Voyager during the occultation was 28.3$^\circ$ \citep{N90}.
This occultation, and those discussed below, are summarized in
Table~\ref{table:equivalent_depths}.

An occultation of the star 28~Sagittarii was observed from Earth in 1989 at
several wavelengths from several observatories when the ring-opening angle was $B_e=25.4^\circ$.  An equivalent depth
at $\lambda=3.1~\mu$m of 3.79$\pm$0.08~km was measured from
observations at the Infrared Telescope Facility (IRTF), and
an equivalent depth of 3.0$\pm$0.1~km was observed at
$\lambda=3.9~\mu$m at Palomar Observatory \citep{N00}. Two separate equivalent depths were obtained at MacDonald
Observatory at $\lambda=2.3~\mu$m: D=2.8$\pm$0.1~km at ingress and
3.6$\pm$0.1~km at egress, highlighting the azimuthal variation in the
F ring \citep{B02}.

In 1995, very close to the solar ring-plane crossing ($B_e=-2.67^\circ$), an occultation of the star
GSC5249-01240 was observed by HST using the Faint Object Spectrograph at wavelengths of 270--740 nm. The F~ring's equivalent depth was measured
to be 7.41$\pm$0.15~km. Simultaneous observations at IRTF at $\lambda=2.3~\mu$m yielded an equivalent depth
of 5.76$\pm$0.06~km \citep{B02}.

The Cassini Visual and Infrared Mapping Spectrometer (VIMS) has
observed 87 stellar occultations by the F ring.  Integrating over a
radial range of 150 km, \citet{H11} find an average equivalent depth
at 3.2~$\mu$m of 6~km, with a standard deviation of 3 km.  A subset of
30 of these occultations were analyzed over a larger radial range of
600 km of radial extent by \citet{F11}, who found an equivalent depth
of 9.99$\pm$2.65 km at 2.92~$\mu$m.

Some observations reveal isolated regions with higher optical
depths. These regions may be composed of larger ring
particles, and some optically thick spikes in occultation profiles may be moonlets.

Occultations observed by the Cassini Ultraviolet Imaging Spectrograph
(UVIS) showed highly variable profiles with several features of higher
optical depth with radial widths of up to 1.4 km, including
sharp-edged features that may be temporary aggregations of smaller
bodies and one opaque feature that could be a moonlet \citep{E08}, and
\citet{A12} detect a wake that is consistent with a moonlet embedded
in the F ring. UVIS observations show that the core usually has a U-shaped
radial profile but sometimes has V- or even W-shaped profiles
\citep{Me10}. A comprehensive analysis of UVIS occultations from
2005--2008 yields an average peak optical depth of 0.43 in the core,
and equivalent widths of the entire F ring ranging from 3.9--50.8 km,
with average of 9.7 km \citep{A12}.

Radio occultations are sensitive to larger ring particles only, and
the parts of the F ring comprising only small dusty particles are not
detected.  Only a compact region with a width of $<1$~km was seen the
Voyager 1 Radio Science Subsystem occultations at $\lambda=$3.6~cm and
13~cm, implying that the larger dusty portion of the F-ring core, not
detected in the radio, is composed of particles with sizes much less
than $\lambda/3\sim1$ cm \citep{M86}.  \citet{S92} measured equivalent
depths from the Voyager profiles of 0.283$\pm$0.035~km and
0.153$\pm$0.066~km at $\lambda$=3.6 and 13~cm, respectively. These
equivalent depths may be a slight underestimate because no measurable
intensity passed through the densest part of the core, a region $\sim
300$~m wide \citep{M86}, but they are small compared to the equivalent
depths of the dusty F-ring component in the infrared.  This region is
absent in 14 of 25 Cassini Radio Science Subsystem (RSS) occultation
profiles at 0.94 and 3.6 cm, and present in only 3 occultations at 13
cm. When present in the RSS occultations, the F ring is 1 km wide with a
peak optical depth of $\sim$0.1 \citep{Ma10}.

Spectral analysis of VIMS occultations show that regions of higher
optical depth tend to have smaller fractions of small particles,
i.e. they are composed of larger particles than the rest of the dusty
ring \citep{H11}.  These large-particle regions are not obvious in
imaging, and it is difficult to characterize their longitudinal extent
because occultations give us information about only an isolated cut
through the F ring. They may be interpreted as a ``discontinuous core''
(see, e.g. \citet{Ma10}) or as isolated regions.  

Though the optical depths of such regions is higher (and moonlets
are, of course, opaque), they tend to be narrow in radial extent. We therefore do
not expect them to make a large contribution to the blocking
efficiency of the F ring, and it is probably the dusty regions
(including the dusty core) that dominate the F-ring equivalent depth.

\subsection{Comparing different measurements of equivalent depth}

Because the F~ring is largely optically thin, the assumed geometry
has little effect on the measured equivalent depth.  Consider a ring
extending radially in the ring plane from $a=r_0$ to $a=r_1$ and
vertically a distance $z_o$ above and below the ring plane.  Let the
ring's opacity be described by some function
$\kappa(a,z)$, while its mass density is $\rho(a,z)$.
 
In an occultation, the intensity of a star, $I_o$, is attenuated to
$I=I_o e^{-\tau_v/\mu}$, where $\tau_v$ is the normal optical depth, and
$\mu=\cos B$, where $B$ is the ring-opening angle to the observer. An
observer directly above the ring plane ($B=90^\circ$) observing an occultation
of a star passing below the rings measures the vertical optical
depth $\tau_v$ as a function of radius, $a$.  The optical depth is defined in 
terms of the absorption coefficient:
\begin{equation}
\tau_v(a)=\int_{-z_o}^{z_o} \kappa(a,z) \rho(a,z) dz.
\label{eq:radial_tau}
\end{equation}
The equivalent depth is
\begin{eqnarray}
{\rm D}&=&\int_{a_0}^{a_1}\tau_v(a) da \\
&=&\int_{a_0}^{a_1} \int_{-z_o}^{z_o} \kappa(a,z) \rho(a,z)~dz~da.
\end{eqnarray}

Consider now an observer of an RPX, who looks radially inward at the ring from a position in the
ring plane ($B=0^\circ$).  Such an observer measures the radial optical
depth, $\tau_r(z)$. Integrating along the line of sight to the star through the F ring,
\begin{equation}
\tau_r(z)=\int_{a_0}^{a_1}\kappa(a,z) \rho(a,z) da.
\label{eq:vertical_tau}
\end{equation}
The vertical analog to equivalent depth is then
\begin{eqnarray}
{\rm D}^\prime&=&\int_{-z_o}^{z_o}\tau_r(z) dz \\
&=&\int_{-z_o}^{z_o}\int_{a_0}^{a_1} \kappa(a,z) \rho(a,z)~da~dz={\rm D}.
\end{eqnarray}
We see that these orthogonal equivalent depth measurements are
identical for an optically-thin ring, as long as the line of sight is
corrected to the radial direction and integration is vertical, or the
line of sight is corrected to normal to the ring plane and the
integration is radial, as the case for these observations
\citep{B02,N00,S92}.

If a ring is optically thick (i.e., $\tau/\mu\gg1$) then when we
attempt to measure the attenuated intensity, we will simply find that
$I=I_o e^{-\tau/\mu}$ approaches 0.  Thus, we can only find the lower
limit for $\tau$, and thus D, based on the limit of sensitivity of the
detector. However, if a ring is optically thin (i.e., $\tau/\mu<1$) in
all of our observations, as is largely true for the F ring in the infrared, we will
always be able to measure some intensity and can calculate the true
value of $\tau$. Whenever we are looking at the same piece of ring, we
will calculate the same D whether we view the ring from above, or the
side, or from any intermediate angle.

\subsection{The inclination of the F~ring}

Using data from several stellar occultations, including one observed near the time of the solar RPX in November 1995, \citet{B02} created a
kinematic model of the F~ring that includes both an eccentricity and
an inclination. We use their Fit \#3, which was computed
using the Saturn pole of \citet{F93}. For reference, the orbital parameters are
listed in Table~\ref{table:bosh}.

\begin{table}
\caption[F-ring orbital parameters.]{F-ring orbital parameters from \citet{B02} 
for the pole of \citet{F93}.}
\label{table:bosh}
\begin{minipage}{\textwidth}
\begin{tabular}{l|l}
  \hline
  $a$          & $140223.7\pm2.0$ km         \\
  $e\times10^3$          & $2.53\pm0.06$               \\
  $\varpi$\footnote{Longitudes are measured from the intersection of
    Saturn's equatorial plane with the Earth's equatorial plane.  The reference epoch is J2000.0=JD 2451545.0.} 
  & $24.1\pm1.6^\circ$          \\
  $\dot\varpi$ & $2.7001\pm0.0004^\circ$/day \\
  $i$          & $0.0064\pm 0.0007^\circ $   \\
  $\Omega_o^a$     & $17.3\pm3.9^\circ$          \\
  $\dot\Omega$ & $-2.6877^\circ$/day         \\
\end{tabular}
\end{minipage}

\end{table}

At its highest point, the F-ring plane rises $a_f \sin i_f = 16$~km
above the main-ring plane, and it dips below the ring plane by the
same amount.  Compared to the $\sim 350$~km radial excursions of the
ring due to its eccentricity, this vertical displacement is very small.  However, the tilt of
the ring becomes important when the ring-opening angle is small enough
that the projected height of the main rings on the sky is comparable
to the F~ring's vertical displacement. For example, at an Earth ring-opening angle of $B_e=0.0050^\circ$, $\sim$4 hours from the exact moment of RPX on 10 August 1995, the
total projected height of the main rings, whose outer radius is $R=137,000$~km, is
only $2 R \sin{B_e} = 24~\rm{km}$, comparable to the F~ring's
vertical displacement.

If the F~ring's line of nodes is neither parallel nor
perpendicular to the line of sight, then the front of the F~ring obscures
one ansa of the main rings more than the other. Conversely, the main
rings block more of the back of the F~ring on one ansa than on the
other.

\section{The photometric model}

\subsection{The Poulet model}

In order to reproduce the observed brightness of the rings of Saturn
in the near-infrared during the 1995 Earth
ring-plane crossings, \citet{P00} constructed a photometric model of
the rings in which the F~ring was the most important contributor to
the brightness of the rings close to the ring-plane crossing.  This
model (hereafter referred to as the Poulet model) was created to
reproduce observations of the dark side of the rings from the
University of Hawaii 2.2-m telescope on 22--23 May 1995, from the
European Southern Observatory (ESO) 3.6-m telescope at La Silla, Chile
on 9--10 August, and of the lit side of the rings from the Pic du Midi
2-m telescope on 12--13 August.  Observations of both the dark and lit
sides of the rings from HST on August 10--11 were also analyzed, but
no attempt was made to reproduce the east-west asymmetry in
brightness.  

In the Poulet model, the F~ring was treated as a uniform uninclined
toroid of rectangular cross-section.  Based on the
shape of profiles extracted from the images of the dark side of the rings
near the RPX, where the light contributed by the main rings was
minimized, \citet{P00} derived an F-ring vertical optical depth of
$\tau_v \sim 0.16 \pm 0.05$ from the HST data ($\lambda$=0.89 $\mu$m), 0.19$\pm$0.05 from the Hawaii
data ($\lambda$=2.2$\mu$m), and 0.27$\pm$0.10 from the ESO data ($\lambda$=2.15$\mu$m).  

\label{sec:p00D}
For the HST data, \citet{P00} found the best fit to the data with a
physical height of the F-ring model of $H=21\pm4$~km, with an
equivalent depth of D=8$\pm$3 km.  This value is larger in general
than the early equivalent depths obtained from occultation data, but
agrees with more recently-measured equivalent depths (see Table~\ref{table:equivalent_depths}).

To fit the phase behavior of the F~ring (albeit over a range of only
$\alpha=3.55$--$5.55^\circ$), the Poulet model employed a dust
component and a macroscopic component.  The fraction of the F-ring
brightness contributed by dust in the model ranged from 0.8--0.9.

\subsection{A new ring model}

\begin{figure} 
\begin{center}
\includegraphics[width=5in]{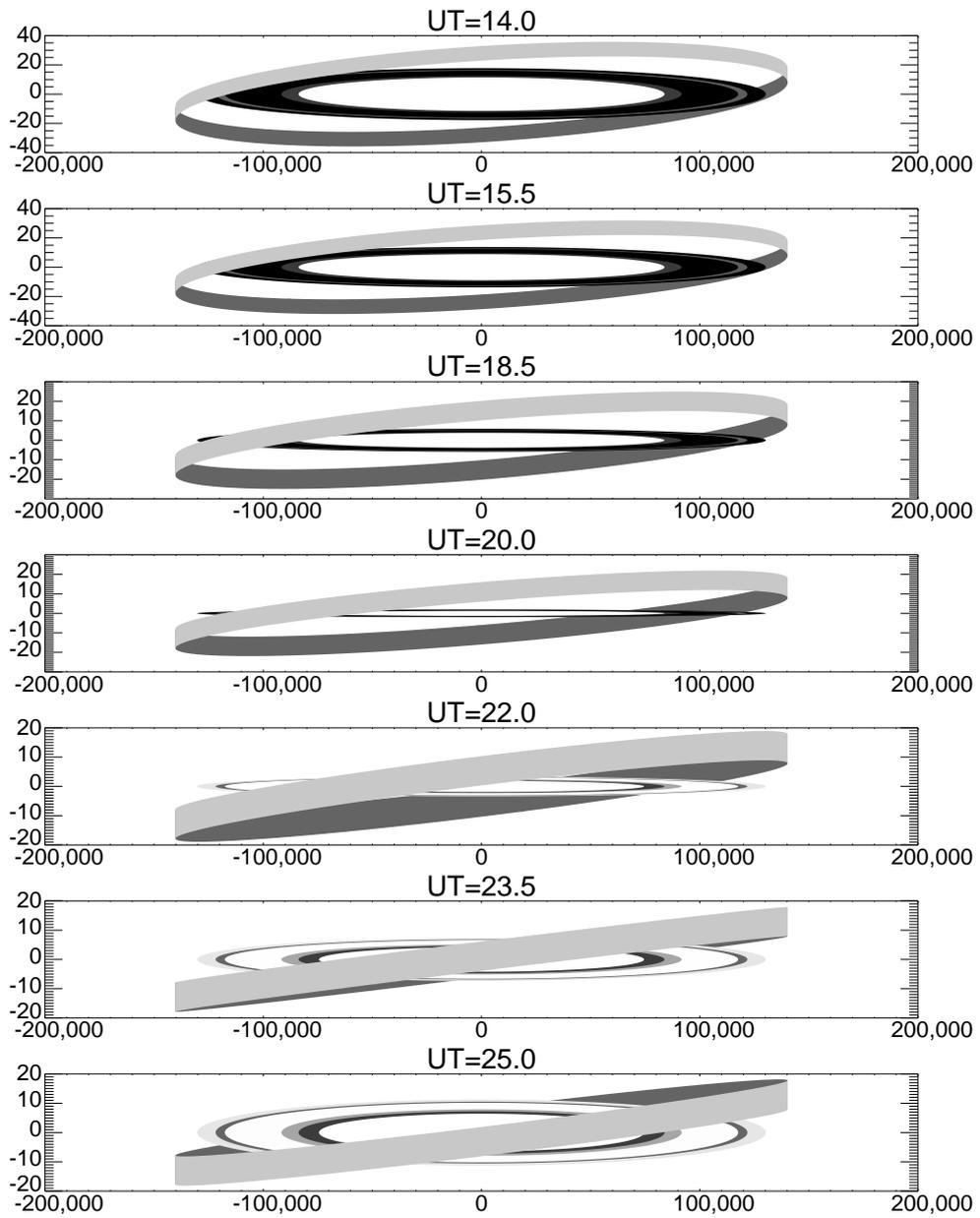}
\end{center}
\caption[Cartoons of the geometry of the rings during the 10 August
1995 RPX.]{Diagrams showing the geometry of the rings at each of the
  HST observation times.  The total vertical thickness of the F~ring
  was arbitrarily set to 10~km in these diagrams.  The Earth crosses the plane
  of the main rings at $\sim$21:00 UT and the plane of the F~ring at
  $\sim$23:30 UT.}
\label{diagram:cartoons}
\end{figure}

While reproducing the observed overall edge-on brightness of the
rings, the uninclined F ring of the Poulet model cannot explain the
difference in brightness between the east and west ansae observed in
HST images of the August 10 RPX seen in Figs.~\ref{plot:hst_correct}
and \ref{plot:hst_wf3}.  In Fig.~\ref{diagram:cartoons}, a very simple
cartoon shows that during this particular ring-plane crossing, the
F~ring's inclination causes it to obscure, and be obscured by, the
main rings by differing amounts on the east ansa and west ansa.  In
particular, immediately after the RPX, the east ansa is more obscured
by the F~ring, then an hour and a half later (as the Earth passes
through the F~ring's plane) the obscuration is similar, and finally
the west ansa is covered by the F~ring.  This qualitative picture
agrees with the asymmetries in brightness observed after the
ring-plane crossing, shown in Fig.~\ref{plot:hst_correct}, where
immediately after the RPX the brightness of the east ansa is less than
that of the west ansa, then the brightnesses are similar, and then the
east ansa is brighter than the west ansa.

In Fig.~\ref{diagram:cartoons}, the F~ring's height was arbitrarily chosen
to be 10~km, but clearly the vertical extent and structure of the F
ring will affect the degree of asymmetry observed at different times.
Thus the brightness asymmetry at ring-plane crossing gives us an opportunity to
probe the properties of the F~ring perpendicular to the
usual radial dimension, which has been well-studied in spacecraft imagery and
stellar occultation data.

Our major improvements over the Poulet model, then, are to include the inclination of the F~ring and to use a more realistic vertical profile of the F~ring's structure. Unlike \citet{P00}, we do not address the particle size in the F~ring, because of the limited phase coverage of the HST data.

The components of our photometric ring model are: the F~ring, divided into a
front half and a back half, and the main rings of Saturn (including
the A, B and C rings, and the Cassini Division).  

The resolution of our model images is 600 km per model pixel in the
horizontal direction (similar to the resolution of the HST WF3 images)
and $\sim$0.03 km in the vertical direction.  The images are stretched
vertically by a factor of $\sim18,500$ in order to correctly model
the F~ring's obscuration of the major features in the main ring (the
narrowest of which is the Cassini Division).  An image size of
500$\times$4000 pixels was chosen to accommodate the vertical extent
of the F~ring during all the model times.  The F-ring model is
constructed as a cylindrical band composed of rectangular cells that
are $\sim \frac{1}{4}$ the size of the pixels in the model image. (For a
detailed discussion of the model resolution, see
\ref{sec:resolution} and Fig.~\ref{diagram:pixels_and_cells}).

For comparison with the HST observations, we model the ring brightness
at the rounded visit times: 14:00, 15:30, 18:30, 20:00, 22:00, 23:30,
and 25:00 UT. There was little systematic variation in the brightness
of the rings in the two WF3 images taken during each dark side visit,
when the rate of change in brightness is small. In the last two visits
on the lit side of the rings, there is only one WF3 image, which fell
very near this rounded time. Only during the 22:00 visit were two
images taken, and there was a significant, systematic change in the
brightness at these two times. However, the brightness of the two
ansae increase at roughly the same rate, so this has little impact on
the asymmetry measurement.  It does lead to a
discrepency in overall brightness between our model profiles and the
observed profiles, as discussed in Sec.~\ref{sec:profiles22}.

The model consists of several image layers at each model time, with the
value of each pixel in each layer representing the pixel's
area-integrated I/F (AIF).  The AIF is then summed over the entire
vertical column of the image layer and divided by the pixel width to
find the VIF.  Finally, the VIF is averaged over 80,000--120,000~km,
the same radial range used to calculate the $\langle$VIF$\rangle$ in
Fig.~\ref{plot:hst_correct} and Table~\ref{table:HST_VIF} and \citet{N96}.

\subsection{The model of the main rings}
\label{sec:main_ring_model}

Table~\ref{table:ring_opening} shows ring-opening angles to the Sun
and Earth, $B_s$ and $B_e$, calculated separately for each visit and interpolated from ephemerides provided by
the Planetary Data System's Rings Node\\ ({\tt
  http://pds-rings.seti.org}). The right ascension and declination of
Saturn, $\alpha_s$ and $\delta_s$, were also obtained from this ephemeris for
use in the calculation of the orientation of the F~ring.

The normal optical depth of the main rings as a function of radial
distance from the center of Saturn was measured in ground-based
observations of the occultation of the star 28~Sgr by Saturn's rings
in 1989 \citep{N00}. We use the optical depth profiles derived from
the data taken with the Lick 1-m Nickel reflector at a wavelength of
0.9 $\mu$m because these observations are of consistent quality and
are at nearly the same wavelength as the HST observations \citep{Fr03}.   We use only the egress data, as the ingress data are noisier
due to thin clouds over the observing site. (This profile is plotted
in Fig.~\ref{plot:tau}.)

To calculate the reflectance of the main rings we also require the
single-scattering albedo of the different regions of the rings and the
phase factor (which we will take to be the same for all parts of the
rings).   Because these photometric parameters, especially the albedo,
are not well-known in the near infrared, we will begin with provisional values,
which will be rescaled in the course of modeling.

\begin{table}
\caption[Single-scattering albedos for the main rings.]{Single-scattering albedos for the main rings based on 0.5-$\mu$m Voyager observations.\\~}
\label{table:main_ring_albedoes}
\begin{minipage}{\textwidth}
\begin{tabular}{c|rrcl}
\hline
       & Inner  & Outer  &  Single-   & \\
 Ring  & Radius & Radius & Scattering &\\
Region & (kkm)  & (kkm)  &   Albedo\footnote{Scaled up
by a factor of 1.2  based on the ring spectra of \citet{K94} at
$\lambda$=0.5 and 0.89~$\mu$m.} & Reference\\
\hline
C ring (inner)   & 74.5  & 83.9  & 0.18 & \citet{C91} \\
C ring (outer)   & 83.9  & 91.7  & 0.31 & \citet{C91} \\
B ring 	  	 & 91.7  & 117.6 & 0.66 & \citet{D89} \\
Cassini Division & 117.6 & 121.9 & 0.38 & \citet{S81} \\
A ring           & 121.9 & 136.9 & 0.6  & \citet{D93} \\
\end{tabular}
\end{minipage}
\end{table}

The nominal single-scattering albedos used in the main-ring model, listed in Table~\ref{table:main_ring_albedoes}, are derived from Voyager clear-filter observations. For our model, which seeks to reproduce the ring brightness at 0.89~$\mu$m, we have scaled these albedos up by a factor of 1.2, based on the full-ring spectrum of \citet{K94}. We neglect small differences in the spectra of the different regions of the rings.

We employ a Callisto-type phase function  \citep{D93} for all regions
of the main rings.  The I/F of the rings is then computed using the
standard single-scattering reflectance formulae from \citet{C60}, as
described in Sec.~\ref{sec:chandrasekhar}.

We have also calculated the contribution of saturnshine to the brightness
of the main rings. Our saturnshine model was tested by reproducing the
observed brightness profile in an image from the Sun ring-plane
crossing on 21 November 1995, including an east-west asymmetry
that was noted by \citet{N96}, which is caused by the relatively large
phase angle at that time. Because the 10--11 August
solar ring-opening angle is much larger than in that November, we find that directly
reflected and transmitted sunlight is much brighter than the
saturnshine, and that saturnshine can safely be neglected in modeling
the data in Figs.~\ref{plot:hst_profiles_e_w} and \ref{plot:hst_wf3}.
The calculation of saturnshine is discussed in detail by \citet{S07},
and a profile of brightness for 21 November 1995 is given in Fig.~\ref{plot:nov_ss}

\subsection{The F-ring model}

The F~ring is constructed as a two-dimensional array of
6000$\times$6000 cells, with a vertical height of $h_f$=60~km,
arranged as an inclined ``ribbon'' in three dimensions. We calculate
the sunlight scattered by the F~ring as well as the amount of light
from the main rings and from the back of the F~ring that is blocked by
the front of the F~ring.

The physical radial width of the F~ring, $\sim 50$~km, being much smaller than the semimajor axis, has very little impact on the photometric model.  Also, the only effect of the eccentricity would be a radial displacement of at most $a_f \cdot e_f=$350~km of the F~ring at the ansa, which is less than the 600-km horizontal resolution of the main-ring model.  We therefore take the radius of the F-ring model, equal to $a_f=140,200$~km \citep{B02}, as a constant for all cells. 


Each cell in the F~ring is assigned a {\em radial} optical depth
according to the formula
\begin{equation}
\tau_r(h)=\tau_o e^{-(h/h_o)^2},
\label{eq:tau_h}
\end{equation} 
where $h$ is the height above the mean F-ring plane, $\tau_o$ is the radial
optical depth at the midplane, and $h_o$ determines the width of the vertical
profile of the optical depth.  $\tau_o$ and $h_o$ are the two major parameters in
our photometric model.

The brightness of the F~ring is also calculated using the single-scattering equations of \citet{C60}, as applied to our model of a vertical ribbon wrapped around Saturn. As the light from the back of the F~ring and from the main rings passes through the front of the F~ring, its intensity will be reduced by a factor of $e^{-\tau_{los}}$, where $\tau_{los}=\tau_r(h)/\cos(\theta)$ is the optical depth along the line of sight. Here, $\theta$ is the angle between the local radial direction and the line of sight to Earth, which varies from -$\pi/2$ to $\pi/2$. Both calculations are described in Section~\ref{sec:f_ring_blocking}.

\subsection{Albedo scaling}
\label{sec:albedo_scaling_summary}

Here we summarize the rescaling of F-ring and main-ring model brightnesses to match the observations.  This process is discussed in more detail in Sec.~\ref{sec:albedo_scaling}.

Because the albedo and phase factor of the F~ring at the wavelength of
this observation is not well-known, we determine it empirically.
Before the ring-plane crossing, the amount of sunlight reflected by
the main rings is small compared to the sunlight reflected from the
F~ring.  When the model is first run, the factor $P(\alpha)\varpi_o$
for the F~ring is set to 1.  The resulting dark-side (pre-RPX) model
brightness is compared to the observed HST $\langle{\rm VIF}\rangle$,
averaged over all dark-side times, and a constant scaling factor $p_f
= P(\alpha)\varpi_o$ is used to re-scale the brightness of the F-ring model so that it matches the observed brightness.
We recompute this factor for each set of parameters $h_o$ and
$\tau_o$, as discussed below.

The albedos of the main rings at 0.89~$\mu$m are also not well-determined.   We run the model with the nominal albedos and phase factors from Table~\ref{table:main_ring_albedoes}, then rescale the brightness of the main rings by a single factor $p_m$  at all times so that the model $\langle{\rm VIF}\rangle$s match the brightness observed in the HST observations {\em after} the ring-plane crossing for each set of model parameters.   This preserves the main ring profile's shape, which depends on the optical depth profile of the rings and the relative albedos of the different regions, while allowing us to correct empirically for uncertainties in the main ring single-scattering albedos in the near infrared.

\subsection{F-ring model parameters}

\begin{figure}
\begin{center}
\includegraphics[width=4in]{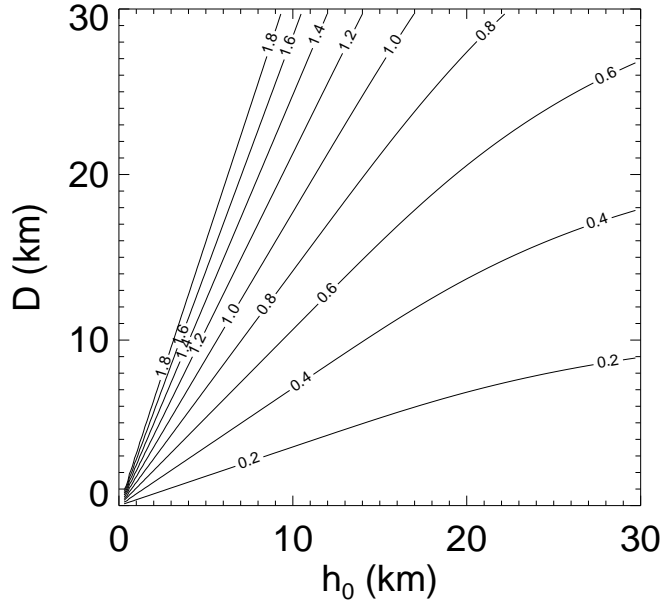}
\end{center}
\caption{A contour plot of $\tau_o$ for a range of values of $h_o$ and equivalent depth.}
\label{plot:tau_contour}
\end{figure}

Besides the albedo scaling factors for the main ring and F~ring, which are calculated in
the course of the model run, there are two fundamental parameters
that are independently varied in this model: $\tau_o$ and $h_o$, which
characterize the F~ring's vertical optical depth profile {\em via}
Eq.~\ref{eq:tau_h}.  The full width (actually, height) at half maximum (FWHM) of the vertical
profile is given by $2\sqrt{\ln 2}~h_o=1.665~h_o$.  In addition to these two parameters, it is also convenient
to consider the model ring's equivalent depth, D.  For our optical
depth profile,
\begin{equation}
{\rm D} =\int^{30~{\rm km}}_{-30~{\rm km}}\tau_r(h)dh 
=\int^{30~{\rm km}}_{-30~{\rm km}}\tau_o e^{-(h/h_o)^2} dh 
       = \sqrt{\pi}~\tau_o~h_o~{\rm erf}\left(\frac{30~{\rm km}}{h_o}\right),
\end{equation}
where erf($x$) is the error function. Because the equivalent depth has been measured in
the F~ring in occultations (see Sec.~\ref{sec:occultations}) we will use
D to represent the total amount of matter in the F~ring,
while $h_o$ describes its vertical distribution, and not refer further
to $\tau_o$.  A contour plot of
the values of $\tau_o$ for a range of D and $h_o$ is shown in
Fig.~\ref{plot:tau_contour}.

\begin{figure}
\begin{center}
\includegraphics[scale=.5]{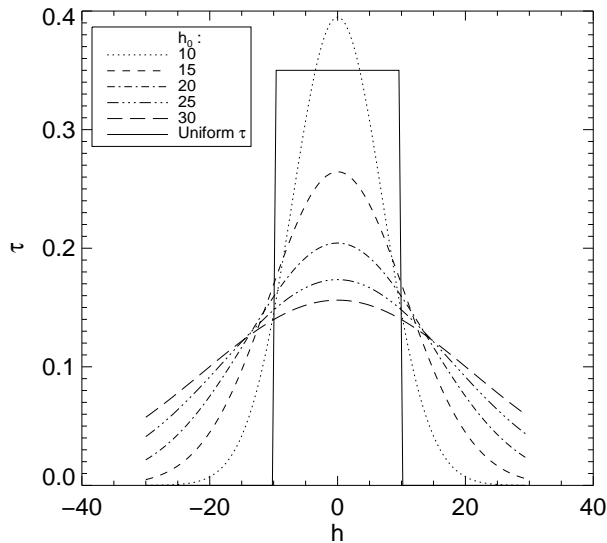}
\end{center}
\caption[Vertical profiles of $\tau_r(h)$ for the F~ring.]{Models of
  the radial optical depth of the F ring as a function of height above
  and below the mean F-ring plane, with varying 
values of $h_o$ and D. The models are cut off at h=-30 and +30~km.  Also
shown is a flat profile with a total height of 21 km, similar to the
F~ring in the Poulet model.  The maximum optical depth of each profile,
including the flat profile, was chosen so that
it has an equivalent depth of 7~km.  }
\label{plot:gaussian}
\end{figure}

Several Gaussians of equal equivalent depth are shown in Fig.~\ref{plot:gaussian}.  A profile with uniform $\tau$, as in the Poulet model, is shown for comparison.  Notice that the profiles are cut off at $|h|$=30~km, as in our F-ring model. The larger $h_o$ is, the flatter the profile.

The choice of model parameters is constrained somewhat by the known
characteristics of the F~ring. As previously discussed, in pre-Cassini
occultation events, the equivalent depth has been measured to be
between $\sim$3 and 8~km at infrared wavelengths.  The best-fitting
Poulet model had an equivalent depth of 8$\pm$3~km. There are no
direct constraints on $h_o$ from previous observations, either Voyager
imaging or stellar occultations, but a plausible assumption is that
the FWHM vertical thickness is not significantly greater than the
radial width of the F~ring's dusty core, or 20--50~km
\citep{L82,S92,B02}, so that we should expect $h_o\lesssim 50 {\rm km}
/ 1.665 = 30$ km.

In summary, the variable parameters in the present model are $p_f$,
the product of the albedo and phase factor of the F~ring; $p_m$, a
scaling factor for the albedos and phase factor of the main rings;
D, the equivalent depth of the F~ring; and $h_o$, a parameter
related to the full height at half maximum of the F~ring.  Whereas
$p_m$ and $p_f$ control the overall brightness of the model and are
adjusted to match the observed brightness of the rings, D and $h_o$
primarily affect the asymmetry and the shape of radial profiles of I/F
(Fig.~\ref{plot:hst_profiles_e_w}).

For each model run, we chose a pair of $h_o$ and D values.  Small
values of $h_o$ have a tendency to increase the modeled brightness
asymmetries, as the F~ring is more concentrated at the core where it
falls across the main rings.  Large values of D, by increasing the
overall amount of blockage by the F~ring, also tend to increase the
model asymmetries unless the value of $h_o$ is also large.

As discussed in Sec.~\ref{sec:albedo_scaling_summary}, $p_f$ and $p_m$ are computed for each model run, i.e. for each pair of $h_o$ and D, in order to bring the overall model brightness into agreement with the observed brightnesses. For small values of D, $p_f$ must be large so that the model F~ring is bright enough to match the observed pre-RPX brightness of the rings, but $p_f$ depends only weakly on $h_o$.  In general, $p_m$ increases as $h_o$ and D increase, because these factors increase the blocking of the F~ring, requiring the main rings to be brighter to reproduce the observed brightness.

\section{Results}

We explored the parameter space of $h_o$ and D to find the combination of these parameters that best fit the observed lit-side asymmetry in the HST dataset.  

\subsection{Best-fitting model}

\begin{figure}
\includegraphics[width=4in]{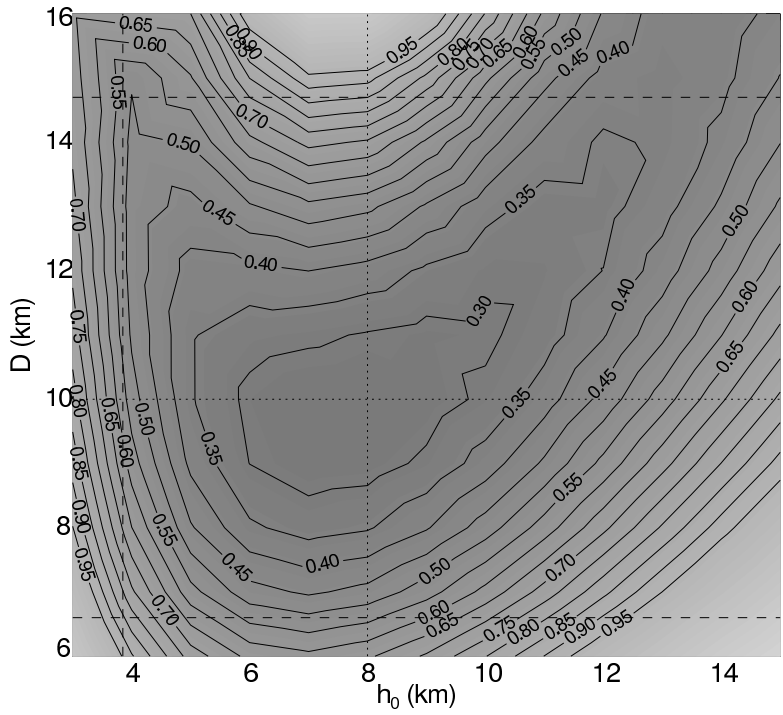}
\caption[The unweighted $\chi^2$.]{Unweighted $\chi^2$ values for a range of values of model
parameters $h_o$ and D.  The dotted lines indicate the position of the
best fit, for $h_o=8$~km and D=10~km.  The dashed lines show the
projection of the 1-$\sigma$ contour, where $\chi^2$=0.5 km, onto the $h_o$
and D axes.  Because this contour does not close at the upper right,
we use the bounds at the lower and left-hand sides to characterize the
uncertainty in the model parameters.}
\label{plot:chi_sq_contour}
\end{figure}

Because we chose the factor $p_f$ for the F~ring to match the observed
$\langle$VIF$\rangle$ on the dark side and $p_m$ for the main rings to
match the observed $\langle$VIF$\rangle$ on the lit side, the overall
model brightness will always be similar to the HST measurements. As
will be discussed in Sec.~\ref{sec:vif_vs_t}, there is poor agreement
between the model and the small pre-RPX asymmetries; whatever
mechanism is responsible for these asymmetries is clearly not
reproduced by the model.   Therefore, we assess each trial model with a chi-square fit to the lit-side asymmetries, where the asymmetries are expressed as the difference in $\langle$VIF$\rangle$ between the east and west ansae at each HST visit.  For each $(h_o,D)$, we compute
an unweighted $\chi^2$,
\begin{equation}
\chi^2=\sum_t (\Delta \langle{\rm VIF} \rangle_t-\Delta \langle{\rm VIF} \rangle_{HST, t})^2,
\end{equation}
where the sum is taken over only the lit-side model times. A contour
plot showing the values of $\chi^2$ that result from models with a
range of $h_o$ and D is shown in Fig.~\ref{plot:chi_sq_contour}.  The
minimum unweighted chi-square is $\chi^2_{min}=0.26$~km$^2$, which is
obtained for $h_o=8$~km and D=10~km, corresponding to $\tau_o=0.7$.

It would be preferable to use a weighted chi-square that takes into account
the standard deviations for each data point, $\sigma_t$:
\begin{equation}
\chi_w^2=\sum_t \left(\frac{\Delta \langle{\rm VIF} \rangle_t-\Delta \langle{\rm VIF} \rangle_{HST, t}}{\sigma_t}\right)^2
\label{eq:weighted_chi_sq}
\end{equation}
\citep{P92}, but \citet{N96} do not quote standard deviations for their 
$\langle$VIF$\rangle$s. Uncertainties for the HST asymmetries based on the
scatter of the HST measurements for each visit are given in
Table~\ref{table:HST_VIF}, but it must be stressed that these include
both WF3 and PC images, whereas $\Delta\langle$VIF$\rangle_{HST}$ is
computed from WF3 images only. Because there are only one or two images
WF3 images per visit, we use the errors derived from the scatter in
the WF3 and PC images together, which are of order 0.5 km, to serve as an
indicator of the general level of uncertainty in the HST measurements.
The uncertainties are dominated by the effects
of superimposed satellites and F-ring clumps, rather than statistical
(i.e., photon) noise.

Lacking rigorous standard deviations for the data, we can assume that a
typical uncertainty in the HST asymmetries is represented
by $\sigma$, so that
\begin{equation}
\chi_w^2=\frac{1}{\sigma^2}\sum_t(\Delta \langle{\rm VIF} \rangle_t-\Delta \langle{\rm VIF} \rangle_{HST,t})^2,
\end{equation}
or 
\begin{equation}
\chi_w^2=\frac{\chi^2}{\sigma^2}.
\label{eq:chi_sq_uw_w}
\end{equation}
The minimum of a weighted chi-square for a reasonable model fit should
be approximately equal to the degrees of freedom.  Because we use the three lit-side data points to determine the two
model parameters, we should have $\chi^2_{w, min}=1$ \citep{P92}.
This suggests that $\sigma=\sqrt{\chi_{min}^2/1}=0.5$~km,
which is in good agreement with the estimated HST uncertainties (on the lit side) in
Table~\ref{table:HST_VIF}. As a test, we also computed a weighted
$\chi^2$ statistic using the errors on $\Delta VIF$ for each HST data
point from Table~\ref{table:HST_VIF} and Eq.~\ref{eq:weighted_chi_sq}.  The results are not very
different from the unweighted $\chi^2$, yielding the same best-fit
parameters, and contours of $\chi_w^2 \sigma^2$ were very similar to
contours of $\chi^2$.  Therefore we take 0.5~km to be a good estimate
for $\sigma$ in order to find the uncertainties in our fitted
parameters.  For a fit with one degree of freedom, the one standard
deviation uncertainty in a single parameter is determined by projecting the region
where $\chi^2_w-\chi^2_{w,min}=1$ onto the axis of that parameter
\citep{P92}.  Again, for one degree of freedom, $\chi^2_{w,min}=1$, so
$\chi^2_w=2$, and using Eq.~\ref{eq:chi_sq_uw_w}, we find that the
uncertainty is defined by the contour where:
\begin{equation}
\chi^2 = 2\sigma^2=0.5~\rm{km}.
\end{equation}

Fig.~\ref{plot:chi_sq_contour} shows how we project the boundaries of
the region where $\chi^2\le0.5$ km onto each parameter's axis.
Unfortunately, solutions with $\chi^2 \sim 0.5$ km exist in a large
region of parameter space for large values of D and $h_o$.  In this
part of the parameter space, the optical depth profile of the F ring
is nearly flat, with the overall height set by our model limits of
$\pm30$~km, and there is little to differentiate between the models as
$h_o$ and D increase.  To estimate the uncertainties of the
parameters, then, we ignore this region and use the better-constrained
low-$h_o$ side of the $\chi^2=0.5$-km contour.  This gives us one-sigma
uncertainties of $\sigma_{h_o}=4$~km and $\sigma_D=4$~km.

For our best-fit model, $p_f=P(\alpha)\varpi_o=0.36$. This parameter
is largely determined by the $\langle$VIF$\rangle$ values measured on
the dark side side of the rings.  \citet{P00} do not quote values for
$P(\alpha)\varpi_o$, but based on their plotted results, we estimate
that the fitted value for their final model was $\sim$0.6. However,
the Poulet model assumed that the back of the F~ring lay entirely
hidden behind the front. Our model shows that, especially before the
ring-plane crossing, significant parts of the back of the F~ring are
unobstructed by the front of the F~ring, effectively doubling the
visible area of the ring. Thus, we are in agreement with the
brightness of the ring in the Poulet
model because our F-ring area is twice as large but our reflectivity is half as large.

We find that for our best-fit model, our main-ring scaling factor is $p_m\sim0.7$, so the main rings are about 70\% as bright as predicted by single scattering using the nominal
photometric parameters.  This does not seem unreasonable, considering the
uncertainty in the
scattering behavior of the main rings at extremely low ring-opening
angles and, to a lesser extent, in the spectrum of the ring in the near infrared.

\subsection{Radial profiles}
\label{sec:profiles}

Profiles extracted from the different component images in the best-fit
photometric model illustrate the time-varying contributions of the main rings and
F~ring to the total profile. These component profiles and total profiles are
shown along with diagrams depicting the geometry of the rings in
Fig.~\ref{plot:cartoons_and_profiles}.  Profiles of VIF($r$) from the
best-fit photometric model ($h_o=8$~km, D=10~km) are over-plotted with
HST profiles in
Fig.~\ref{plot:hst_model_profiles}.

\begin{figure} [b]
\caption[Model profiles of VIF($r$) and diagrams the ring geometry.]{Model profiles of VIF($r$) and corresponding diagrams of the ring geometry for each of the HST observations. The F~ring
is depicted in the diagram as a 10~km-high ``ribbon,'' but its
full height in the photometric model is 60~km.  The total brightness,
the sum of the other components shown, is plotted as a solid gray line.  The dash-dot line shows the
brightness of the main rings (MR). A dashed line gives the
brightness of the back of the F~ring, with light blocked by the main
rings removed (BF). The dotted line is the sunlight
reflected from the front of the F~ring (FF). The solid black line
shows the blocking by the front of the F~ring of the main rings
and the back of the F~ring (FBF+FBM), and has negative values.  The
profiles were created with our best-fitting F-ring model parameters,
$h_o$=8~km and D=10~km.  Note the change in the scale of the y axis in the last three panels.}
\label{plot:cartoons_and_profiles}
\end{figure}
\begin{figure}
\includegraphics[width=\textwidth]{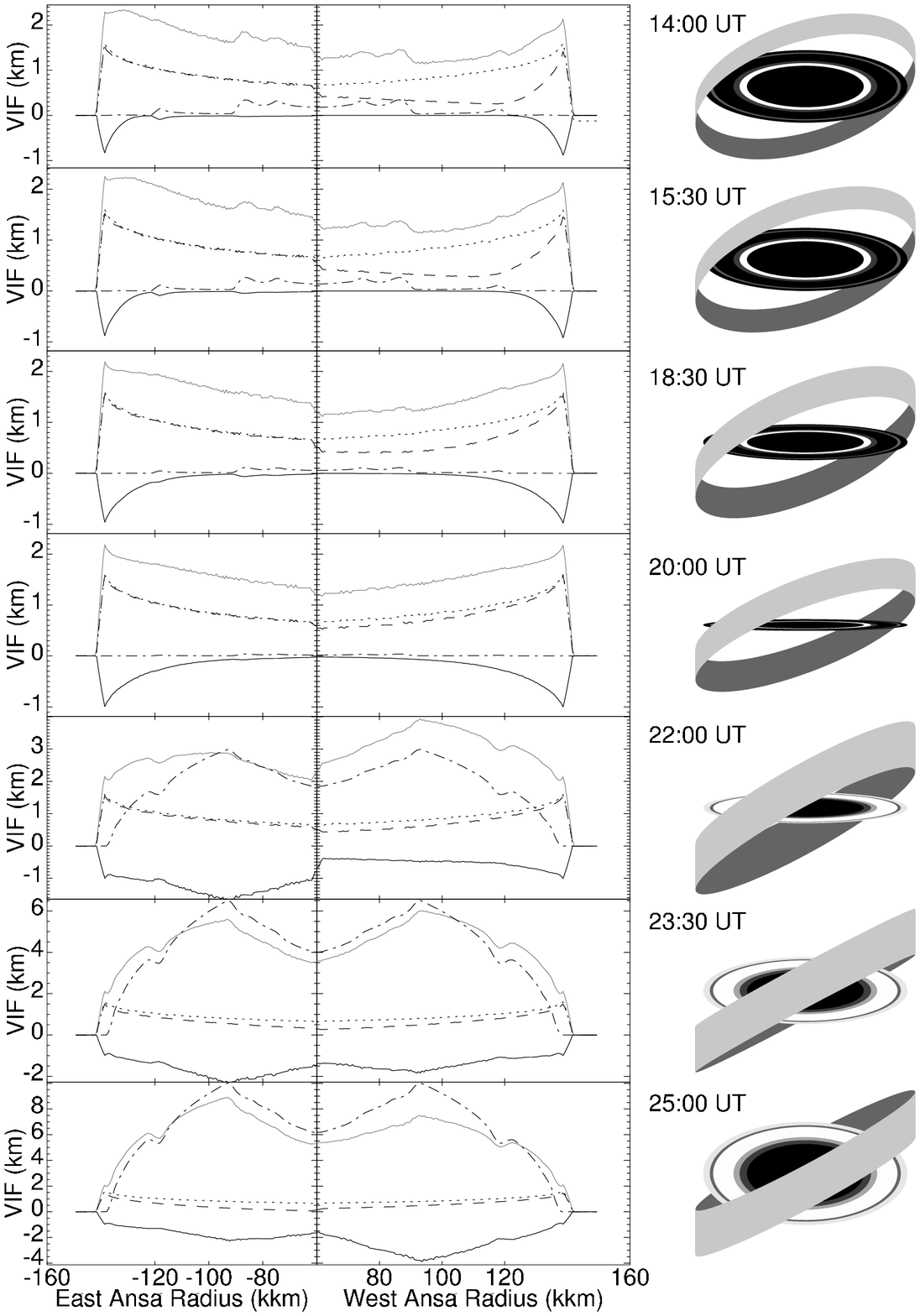}
\end{figure}

\begin{figure}
\includegraphics[height=6.5in]{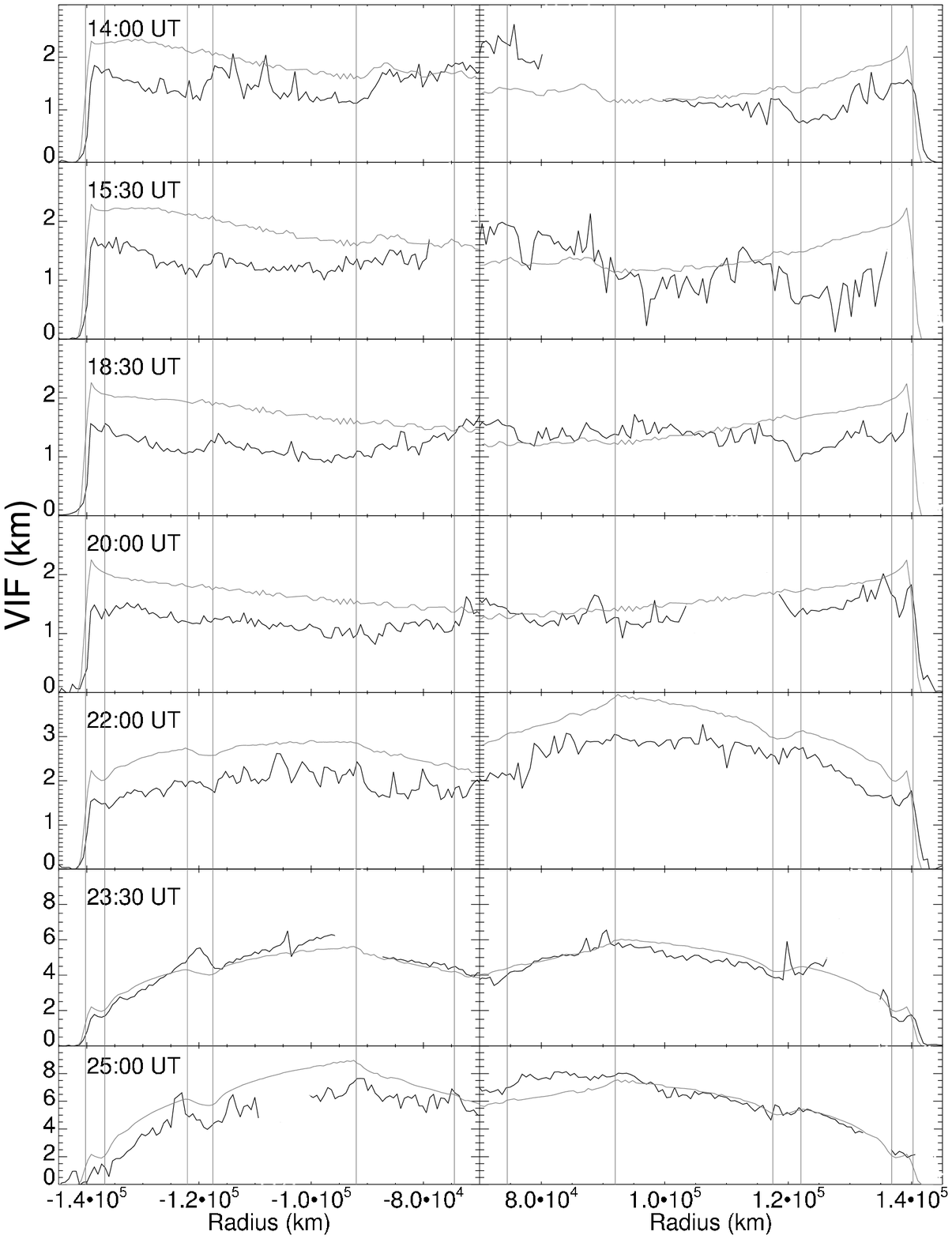}
\caption[Profiles of VIF($r$) from the HST data and the model.]{Profiles of VIF($r$) from the HST data (black line) and from the 
model (gray line) for the best-fit parameters D=10~km, $h_o$=8~km.  The
vertical lines indicate the boundaries of the different ring regions
as in Fig.~\ref{plot:hst_profiles_e_w}.  Unlike 
Fig.~\ref{plot:hst_profiles_e_w}, in this figure we have removed the
regions in the HST profiles that are contaminated by light from
satellites. }
\label{plot:hst_model_profiles}
\end{figure}

\subsubsection{Dark-side profiles}
\label{sec:dark-side_profiles}

The model profiles are a fair match to the generally flat shape of the
HST profiles on the dark side of the rings (i.e., 14:00--20:00~UT).
The sunlight reflected from the F~ring dominates the model brightness,
contributing VIF$\sim$1.5~km for each ansa.

On both ansae, the model profiles slope upward toward the ansa with a
shape typical of a narrow, optically thin ring.  However, the HST
profiles are flatter at the ansae.   For an
optically thick narrow ring, the I/F would not increase toward the ansa
because $\tau/\mu$ would always be $\gg1$, so, using the single-scattering
expression for reflected light from \citet{C60},  in the optically thick limit,
\begin{equation}
{\rm I/F} \approx \frac{1}{4} P(\alpha) \varpi_o \frac{\mu_o}{\mu+\mu_o}  \approx  \frac{1}{4} P(\alpha) \varpi_o \frac{1}{\mu/\mu_o+1} \approx \frac{1}{8} P(\alpha) \varpi_o,
\label{eq:IF_optically_thick}
\end{equation}
leads to a profile that is flat all the way out to the ansa.  The
flatter HST profiles could therefore be consistent
with an optically thick layer of larger particles in the F-ring
plane and a more diffuse
envelope, suggesting a possible direction for future models.  

Prior to RPX, the main rings are dark, and most of the blocking by the front of the F~ring,
represented by the solid line of negative values in
Fig.~\ref{plot:cartoons_and_profiles}, is of the reflected sunlight
from the back of the F~ring.  This blocking is symmetric east-to-west,
and increases as the optical depth along the line of sight increases
toward each ansa of the F~ring.

The small asymmetry in the model arises from the blocking of
the back of the west ansa of the F~ring by the main rings, as can be
seen by comparing the east and west dashed profiles in
Fig.~\ref{plot:cartoons_and_profiles}. This blocking decreases as the
ring-opening angle decreases, decreasing the projected area of the
main rings and is essentially zero at 20:00~UT, at which point, the
HST profiles are also symmetrical.

In the HST images, the asymmetry between the ansae is smaller than in
the model and the west ansa is slightly brighter than the east ansa
at 18:30 and 20:00~UT (cf. Fig.~\ref{plot:hst_wf3}), opposite to the
sense of the asymmetry caused by blocking of the F~ring by the main
rings.  The model does not reproduce this observed asymmetry, and the
HST profiles do not clearly show the asymmetry predicted by the model
due to the blocking of the back of the F~ring on the west ansa by the main rings.  This
blocking, which amounts to $\lesssim0.5$~km of the total VIF, may
simply be lost in the ``noise'' of the actual F~ring's well-known azimuthal variation.

In the model profiles at 14:00 and 15:30 UT, the transmitted light from
the C~ring contributes a VIF of $\sim$0.2~km and is visible interior
to 90,000 km, a feature visible in the HST profiles as well.  The
C~ring's brightness decreases with the ring-opening angle, and it is not
visible in either the HST nor the model profiles for 18:30 and
20:00~UT.

As can be observed in the diagram of the rings, before RPX, the
Cassini Division is obscured by the F~ring on the east ansa. (Note
that, whereas the F~ring in the photometric model is 60~km in total
height, the F~ring in the diagram is depicted with a height of only
10~km so that the main rings can be more plainly seen.  This narrow
band also serves to indicate the position of the densest part of the
model F~ring.)  In the model profiles, the Cassini Division is barely
discernible on the west ansa with a VIF$\sim$0.1 km near 120,000 km,
and there is almost no trace of it on the east ansa where the F~ring
blocks most of the light from the Cassini Division.  In most of the
HST dark-side profiles there seems to be a small feature with an
amplitude of $~\sim0.3$ km at the radius of the Cassini Division. It
is more prominent on the west ansa.  However, the level of noise in the HST profiles makes a positive identification of this feature uncertain.

There are several localized features in the HST dark-side profiles
which are not accounted for by the model.  Many of these features are
present in all the HST profiles that were combined to produce the
composite profiles and are not due to an anomaly in a single
image. (See, e.g., the two bright features of ~0.5 km just inside the
Cassini Division and in the C ring in the east ansa profile at 14:00
UT in Fig.~\ref{plot:hst_model_profiles}.)  They are also not due to
satellites, which have been largely removed by
median-filtering. (Portions of the profiles where satellites were not
removed effectively have been blanked out in
Fig.~\ref{plot:hst_model_profiles}.) Most of these features are
probably a result of azimuthal structure (e.g., clumpiness) in the
F~ring, which is not included in our model. We have found it difficult
to track these features with the orbital motion of the F~ring in this
dataset due to the $\gtrsim 1$~hr gap between visits and the known
moons of Saturn that frequently obscure key regions of the profiles,
and so are unable to confirm that they are part of the F~ring.

\subsubsection{Lit-side profiles}

The asymmetry in the model brightness after the RPX results arises almost entirely from the differences in the blocking of the sunlit side of the main rings by the front of the F~ring.

At 22:00~UT, just $\sim$1~hour after the main-ring-plane crossing,
our component profiles show that sunlight reflected from the main rings is already comparable in
brightness to that reflected from the F~ring, and as $B_e$ increases further,
so does the VIF; by 23:30~UT the light from the main rings dominates
the system's brightness (Fig.~\ref{plot:cartoons_and_profiles}).  Notice that the F~ring's brightness is
almost independent of the ring-opening angle.

At 22:00~UT, the west ansa is brighter than the east ansa in model
profiles, because the dense central region of the F-ring model lies
directly in front of the east ansa but falls north of the west ansa (see
Fig.~\ref{plot:cartoons_and_profiles}).  The observed
asymmetry and the shape of the HST profiles at this time are reproduced well by the
model. The
maximum in VIF at the inner boundary of the B~ring, a slight dip at
the Cassini Division, and the small peak at the radius of the F~ring
on both ansae are seen in both HST and model profiles.  The east ansa
profile is flatter than the west ansa because of
the blocking of the B-ring region at this time (see
Fig.~\ref{plot:cartoons_and_profiles}). \label{sec:profiles22} The HST profiles are
$\sim$0.5~km lower than the predicted values at this time but this is
mostly because the bulk of the HST images were taken a few minutes
before 22:00~UT, the time chosen to represent this HST visit in the
model, and the main rings are opening rapidly (see
Fig.~\ref{plot:hst_correct}).

At 23:30~UT, near the F-ring-plane crossing, the model F~ring obscures
both ansae to a similar extent, leading to little if any asymmetry in ring
brightness.  The shape of the model profiles is also
quite a good match to the HST profiles.

At 25:00~UT, the F~ring still partially blocks the east ansa, but blocks much more of the west ansa. Considering the noisiness of the HST east ansa profile, which is made from the mean of only two profiles and contaminated by a bright satellite, the agreement in the general shape of the profile is fair. It is very good on the west ansa, except in the inner C ring, where the HST VIF is $\sim 1$~km greater than the model.
We suspect that this is a clump in the F~ring itself, but due to the
noise in the HST profiles and the presence of contaminating
satellites, we were unable to definitively identify this feature at
any other times, and thus we cannot confirm that it is an F-ring feature.

\subsection{$\langle$VIF$\rangle$ vs. t}
\label{sec:vif_vs_t}

\begin{figure}
\begin{center}
\includegraphics[width=5in]{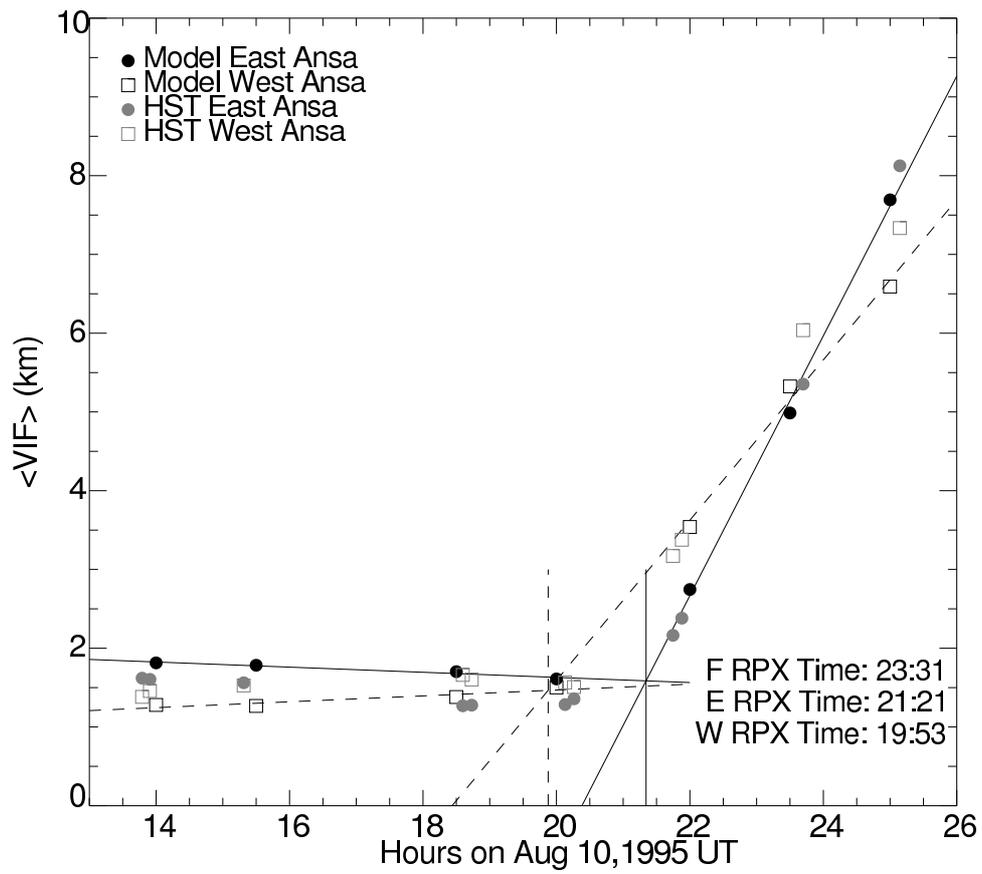}
\end{center}
\caption[The best-fit model $\langle$VIF$\rangle$ plotted vs. time.]{$\langle$VIF$\rangle$ as a function of time, showing the
model results (black) and the HST WF3 data (gray).  The intersections of 
separate linear fits to the model 
brightness before and after the RPX are used to calculate a RPX time
for each ansa. The intersection of the east and west ansa fits
on the lit side gives the FRPX time.}
\label{plot:model_vif_vs_t}
\end{figure}

While the comparison of the model and HST profiles of VIF(r) discussed
in Sec~\ref{sec:profiles} demonstrates that the model reproduces the
HST data in a qualitative way, our model is actually
constrained by the brightness asymmetry between ansae in
the HST observations.  Fig.~\ref{plot:model_vif_vs_t} shows the
$\langle$VIF$\rangle$ of our best-fitting model along with the
$\langle$VIF$\rangle$ measured from each ansa of the WF3 HST images. 

Before the ring-plane crossing, the average model brightness of all dark-side points (including the east and west ansae) matches the average HST brightness because this is how we scaled the albedo of the model F~ring.  As discussed in Sec.~\ref{sec:dark-side_profiles}, it is clear that the asymmetries in the model do not match the asymmetries in the HST data on the dark side. 

On the lit side, however, the model provides a good match to the HST
data.  At 22:00~UT, the west ansa is brighter than the east ansa in
both the model and the data and by about the same amount.  At 23:30~UT,
the west ansa is still brighter, but the amplitude of the asymmetry has
greatly decreased.  By 25:00~UT, the east ansa is brighter than the west ansa,
and the asymmetry is large once again.

To compare to \citet{N96}, we do a linear fit to the model predictions
for each ansa, fitting the points before and after the RPX separately.
The intersection of these fits gives a model RPX time for each ansa.
We find a ring-plane-crossing time of 19:53~UT for the west ansa and
21:21~UT for the east ansa.  As discussed previously, from the WF3 HST
data, we find a west-ansa crossing time of 20:25$\pm$10~min and an
east-ansa crossing time of 21:13$\pm$3~min.  The model RPX time on
the west ansa differs significantly from the HST time, in part due to
the failure of the model to reproduce the change in brightness with
time of the rings before the ring-plane crossing, and in part due to
the low slope of the fit to the lit-side west-ansa model VIFs. 

We can also compute the intersection of the east and west ansa {\em
  lit-side} linear fits to find the time for the F-ring-plane crossing
(FRPX).  This result is more robust, because it does not involve the
dark-side model prediction, and, because it depends on the orientation
of the F~ring, allows us to assess the F-ring orbital model that was
used to create our photometric model. From the HST data, if we fit
both PC and WF3 data, as in Fig.~\ref{plot:hst_correct}, the FRPX time is
23:24~UT.  Using just the WF3 data as in Fig.~\ref{plot:hst_wf3}, we compute an FRPX time of 24:09~UT (which is actually 00:09~UT on 11 August).  We take the difference of 45~min as an indication of the experimental uncertainty in the FRPX time.

In the model results, the time of the F-ring-plane crossing shows
little dependence on the choice of photometric parameters of the
F~ring. It does depend on the orientation of the F-ring plane, which
is determined by the inclination and the ascending node. To find the
sensitivity of the model results to small changes in the orbital elements
of the F~ring, we made several models with values for the inclination
and the longitude of the ascending node increased and decreased by the
standard errors given by \citet{B02}.  All these runs were conducted
with $h_o=10$~km and D=5~km.

For nominal values of the inclination and the longitude of the
ascending node ($i=0.0064^\circ$ and $\Omega_o=17.3^\circ$) for this
model, we find a predicted FRPX time of 23:29~UT. Increasing the
longitude of the ascending node, which moves the node to the west on
the sky, causes an earlier FRPX.  For the nominal inclination and $\Omega_o=13.4^\circ$, the FRPX time is
24:06~UT on 10 August 1995, and for $\Omega_o=21.2^\circ$ the FRPX
time is 23:21~UT. Increasing the inclination of the F~ring has the
effect of delaying the FRPX time.  With the nominal value for the
longitude of the ascending node at epoch and an inclination $i=0.0057^\circ,$
the FRPX time was computed to be 23:20~UT, and for $i=0.0071^\circ$,
the FRPX time was 24:09~UT.

Thus, varying the inclination and longitude of the node within the
quoted errors of the \citet{B02} orbit can change the time of the FRPX
calculated from model results by $\pm 25$~min, which is comparable to
the uncertainty in the FRPX time that is measured from linear fits to
the HST data. Because the F-ring blocking affects the model brightness
before and after RPX, varying $i$ and $\Omega_o$ also affects the
model asymmetries and changes the main ring crossing times.  However, this change is less than 2~min. The changes in the asymmetries resulted in changes in $\chi^2$ of up to 17\%, much less than the 1-$\sigma$ level of expected statistical variation.

Based on this experiment, we are confident in using the best-fit F-ring orbit of
\citet{B02} to determine the geometry for our model.  Our model
results show that a change in $i$ or $\Omega_o$ much greater than the
stated 1-$\sigma$ uncertainties would lead to a significantly poorer fit to the HST data,
but we do not have enough sensitivity to the node or inclination in
these data to improve on the fit of \citet{B02}.

\section{Conclusions}

Our photometric model demonstrates that the inclination of the F~ring
provides a plausible explanation for the asymmetry in brightness
between the east and west ansae of the rings of Saturn within a few
hours after the Earth ring-plane crossing of 10 August 1995, as
originally suggested by \citet{N99}.

\subsection{Profiles of Brightness vs. Distance}

Despite its simplicity, the model also does a fair job of reproducing the
general shape of the HST profiles. Both observed and model profiles of
the dark side of the rings are flat, with a slight increase in
brightness at the ansa.  The shape of the lit-side model profiles are
a particularly good match to the HST data.  They show the observed features of the main ring profile
clearly, including a small brightening at the location of the Cassini Division,  and show how the brightness of the main rings increases with the ring-opening angle.  The model profiles also show how blocking of the main rings by the front of the F ring flattens the shape of the HST profiles.

\subsection{Equivalent Depth}

The best-fit equivalent depth of this model, D=10$\pm$4~km, is in
agreement with the equivalent depth of 8$\pm$3~km of \citet{P00}, but
both of these are larger than the equivalent depths derived from
profiles of optical depth of the F~ring obtained from occultations
observed from Voyager and from Earth in the 1980s.  However, as
shown in Table.~\ref{table:equivalent_depths}, the equivalent depth of
that we measure for the F ring may fit into a trend
of increasing equivalent depth observed between the Voyager and
Cassini epochs.

There are also other hints that the F~ring's behavior is also not
uniform over decadal timescales.  Averaging the equivalent depths
reported for 87 Cassini VIMS stellar occultations yields a mean equivalent
depth of 6$\pm$3~km at 3.2~$\mu$m \citep{H11}, which agrees with the
results the present model and \citet{P00}.  Stellar occultations
observed with Cassini UVIS yield equivalent depths between 4 and an
extreme upper limit of 51~km at different longitudes \citep{A12}.  By
comparing the brightness of the F~ring in Cassini and Voyager images,
and optical depth profiles from occultations from both missions,
\citet{F11} found that the brightness and
equivalent depth of the F ring has increased by a factor of ~2--3, since the Voyager epoch. 

It is likely that this change is due to small particles being freed
from the surface of larger parent bodies due to collisions or
gravitational interactions between objects inside and near the F ring,
but the exact mechanism is unknown.  \citet{F11} reject close
encounters between Prometheus and the core at Prometheus' apoapse as
the sole cause of the brightening, because Prometheus' distance from
the F-ring core has varied significantly during the Cassini mission,
will the F ring's brightness has stayed fairly constant.

\subsection{Azimuthal Variation}

The mechanism
which produces an asymmetry in ring brightness on the dark side of the
rings in our model, i.e. blocking of the back of the F~ring by the main
ring, does not account for the asymmetry in the brightness observed in
the HST images prior to RPX.  Instead this is probably due to the significant longitudinal
variations exhibited by the F~ring.

It is well-known that the characteristics of the F~ring are not
uniform longitudinally due to complex dynamical interactions with
Prometheus \citep{K90,M08} and collisions with smaller moonlets
\citep{K90, B10}, and \citet{S09} suggested activity related to clumps
can affect the F ring's appearance on short timescales.  Cassini
occultations at a variety of longitudes distributed around the F~ring
find a range of equivalent depths measured at different longitudes
demonstrates that the azimuthal variability of the F~ring is more than
enough to cause the small but noticeable asymmetry observed in the HST
data before the RPX \citep{F11,A12}.

Future photometric models of the F~ring can take advantage of far
denser coverage in time and longitude from the Cassini spacecraft, and
may be able to address the variation in the F ring to create a more
complete picture of its azimuthal structure.  As the 1995 Earth RPX
was used to probe the F ring's vertical structure, ring-plane
crossings by Cassini and the 2009 solar ring-plane crossing observed
by Cassini have similar potential to improve our understanding of the
vertical extent and shape of the F ring even as it varies in longitude
and in time.  In future publications, we hope to apply our model to
Cassini VIMS RPX observations.

\appendix
\section{A detailed description of the photometric model}

This appendix describes the computer model which was used to simulate the brightness of the ring system during the 1995 August ring-plane crossing.

The ring model is constructed from arrays of pixels, in the case of
the main rings, or ``cells,'' in the case of the F~ring.  For each
pixel or cell, we assign an optical depth, $\tau$, a single-scattering
albedo, $\varpi_o$, a phase factor $P(\alpha)$ and coordinates in
space and on the sky. The F ring is separated into a front portion and
a back portion.  We then create component image layers by computing
the transmitted and/or reflected sunlight for each part of the ring
model, as well as the transmission of light from the model rings that
lie behind, as appropriate. We construct profiles of VIF vs. radius
from Saturn's center for each component, and then sum the profiles to produce the final model profile.

\subsection{Image layers and model resolution}

Each image is an array $500\times4000$ pixels in size.  Each pixel is assigned $u$ and
$v$ coordinates, with (0,0) being the center of Saturn. The $\hat u$
axis points westward along the intersection of the equatorial plane of
Saturn and the sky plane, and the $\hat v$ axis points along the north
pole of Saturn projected onto the sky.  
The value of each pixel in an
image layer is the area-integrated I/F (AIF) for that position on the
sky.

\label {sec:resolution}
We do not vary the model resolution for the different model times; the
pixel and cell sizes were chosen so as to be appropriate for both
large and small ring-opening angles. 
The horizontal size of the u-v image is 300,000~km, or 500 pixels,
large enough to accommodate the rings' full diameter with a horizontal pixel size of $\sigma_u=600$~km, comparable to the HST resolution of $\sim$650~km.

We require sufficient vertical resolution to model the F~ring's
obscuration of major main-ring features. 
The narrowest important ring
feature, the Cassini Division, has a width of $\Delta a$=4537~km.  Its
projected height on the sky is smallest at the limb of Saturn, where it is 0.07~km for the smallest $|B_e|$ used in our model, at 20:00 UT.  We therefore chose
$\sigma_v=0.0325$~km/pix, and 4000 pixels are required in order to accommodate the extent of the F-ring model at the largest ring-opening angle for the HST dataset, when it has a total height of 130~km on the sky.

We chose the horizontal size of the F-ring cells to be approximately
$\frac{1}{4}$ the size of the main-ring pixels. This results in 6000
cells azimuthally around the F~ring with a horizontal size of $2 \pi
a_f/6000=$146.6~km/cell.  We chose 6000 cells in the vertical
direction, since 6000$\times$6000 was the largest array size for which
the modeling computer had sufficient memory.  This gives us a vertical
cell size of $0.01$~km, or about $\frac{1}{3}\sigma_v$.

\begin{figure}
\begin{center}
\includegraphics[width=3in]{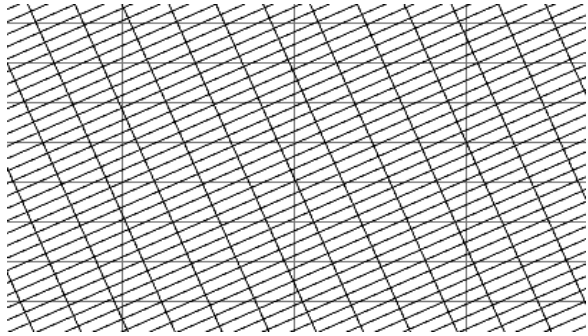}
\end{center}
\caption[Relative size and orientation of main-ring pixels and F-ring cells.]{The relative size and orientation of the main ring pixels and
F-ring cells. The main ring pixels (horizontal grid) are $\sigma_u$=600~km wide
by $\sigma_v=$0.0325~km high.  The F-ring cells are 146.6~km wide by 0.01~km
tall, and are tilted due to the inclination of the F~ring.  This
picture shows a region of the u-v plane $\sim$2000~km wide by
$\sim0.3$~km tall, and the vertical scale in this diagram is
exaggerated by a factor of 400.  A full u-v image is 300,000~km by
130~km.}
\label{diagram:pixels_and_cells}
\end{figure}

Fig.~\ref{diagram:pixels_and_cells} shows the relative sizes and orientations
of the main-ring pixels and the F-ring cells as they appear in the model.

The five image layers that make up the model are:
\begin{enumerate} 
\item The sunlight reflected from the back of
the F~ring minus that which is blocked by the main rings (henceforth
labeled BF).
\item The sunlight reflected or transmitted by the main rings
(MR).
\item The light from the back of the F~ring that is blocked by the
front of the F~ring (FBF).
\item The light from the main rings that is
blocked by the front of the F~ring (FBM).
\item The sunlight reflected
from the front of the F~ring (FF).  
\end{enumerate}

We compute the AIF of each pixel in each layer. The AIF values of the pixels of the blocked-light images (FBF and FBM) are negative, indicating light that is absorbed or scattered by the F~ring so that it does not reach the observer.

Theoretically, a total model image could be constructed by summing the
image layers. However, because the rings are not
vertically resolved in the HST images, there is no reason to produce a
final image of the model.  Instead, as each layer is
computed, a profile of vertically-integrated I/F is extracted, and once it is no longer needed, the image is deleted to free up memory.  The VIF profile is given by:
\begin{eqnarray}
{\rm VIF}(u) &=&\frac{1}{\sigma_u} \sum_v {\rm I/F_{pix}}(u,v) \cdot
A_{pix} \\
                 &=&\frac{1}{\sigma_u} \sum_v {\rm AIF_{pix}}(u,v) \\
                 &=&\sigma_v\sum_v {\rm I/F_{pix}}(u,v),
\end{eqnarray}
where I/F$_{pix}(u,v)$ is the reflectance computed for the pixel at
coordinates $(u,v)$ in the image, and $A_{pix}=\sigma_u\sigma_v$ is
the projected area of the pixel on the sky.  The $u$ coordinate is
equivalent to $r$, the horizontal distance from the center of Saturn,
measured in the plane of the sky, giving us a profile, VIF($r$).  A
profile representing the total light scattered from the rings is
obtained by summing the profiles of the components, some of which are negative.  This is referred to as the total profile, VIF$_{tot}(r)$.

The total VIF profile can then be integrated radially to compute the radially
averaged, vertically-integrated I/F:
\begin{eqnarray}
\langle {\rm VIF} \rangle &=& \frac{1}{120,000 - 80,000~{\rm km}}\int^{120,000~{\rm km}}_{80,000~{\rm km}}{\rm VIF} \cdot dr\\
                         &=&  \frac{\sigma_u}{40,000~{\rm km}} \sum_{u} {\rm VIF}(u).
\end{eqnarray}
We average over the same range that was used to compute $\langle$VIF$\rangle$ by \citet{N96}.  Just as the total model profile is the sum of the component profiles, the total $\langle$VIF$\rangle$ of the ring system is then the sum of the $\langle$VIF$\rangle$s of the components.

\subsection{The Main Ring Model}

\label{sec:chandrasekhar}
The reflectance of the main rings in transmitted sunlight (before the
RPX) or reflected sunlight (after the RPX) is computed using the
single-scattering functions of \citet{C60} for a uniform
plane-parallel medium. When the Earth is on the dark side
of the rings, we use the formula for transmitted light,
\begin{equation}
{\rm I/F} = \frac{1}{4} P(\alpha) \varpi_o \frac{\mu_o}{\mu-\mu_o} 
\left( e^{-\tau/\mu} -e^{-\tau/\mu_o}\right).
\label{eq:IF_trans}
\end{equation}
and when the Earth is on the lit side of the rings, we use the formula for reflected light,
\begin{equation}
{\rm I/F}= \frac{1}{4} P(\alpha) \varpi_o \frac{\mu_o}{\mu+\mu_o}\left(1-e^{-\tau(1/\mu+ 1/\mu_o)}\right),
\label{eq:IF_ref}
\end{equation}
In these expressions, $\mu=\sin|B_e|$ and $\mu_o=\sin|B_s|$, where $B_e$
and $B_s$ are listed in Table~\ref{table:ring_opening}.

The area-integrated I/F of the pixel is then:
\begin{equation}
\rm{AIF}=\rm{I/F}_{pix} \sigma_u \sigma_v.
\end{equation}

The center of the planet is located at the center of the
image.  Using u-v coordinates, the distance of each pixel from
the planet's center in the ring plane is computed: $R= \sqrt{ u^2+ (v
  / \sin{B_e})^2 }$.  An albedo is assigned to each pixel of the
main-ring array based on its radius, $R$, using the values in
Table~\ref{table:main_ring_albedoes}.

\begin{figure}
\includegraphics[width=\textwidth]{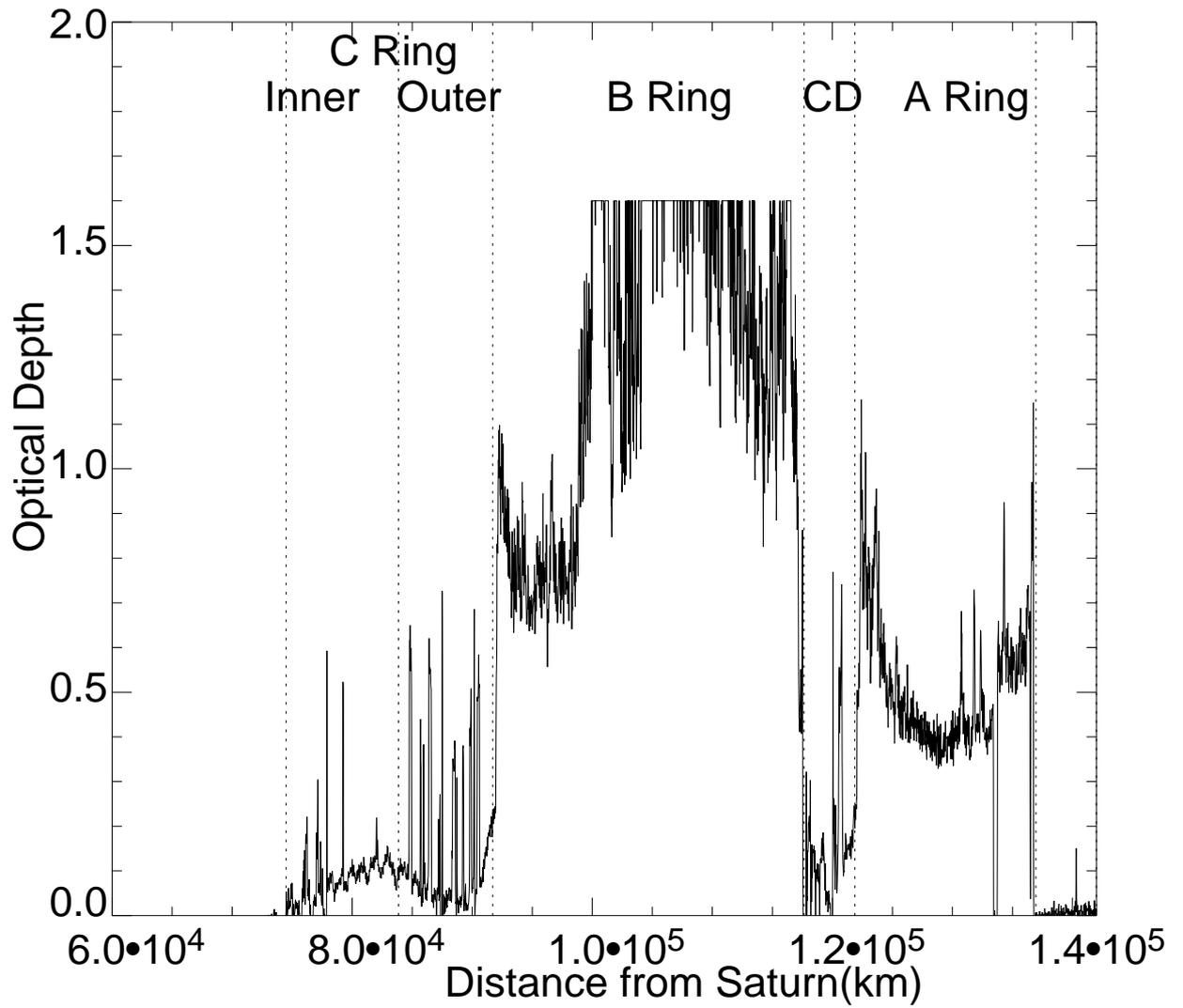}
\caption[$\tau(R)$ for the main rings.]{The profile of optical depths used for the main-ring model,
taken from an observation of the occultation of the star 28~Sgr by the
rings \citep{N00}.  This plot is cut off at $\tau=1.6$, which is
the maximum detectable optical depth in the Lick egress data. The
affected regions in the B~ring are all optically thick, so the exact
value of $\tau$ is not important for calculating the I/F near
RPX.  }
\label{plot:tau}
\end{figure}

Optical depths are assigned in a similar way, based on the Lick
optical depth profile shown in Fig.~\ref{plot:tau}.  The majority of the
transmitted  light in the model passes through the Cassini Division
and the C Ring (as seen in Fig.~\ref{plot:cartoons_and_profiles}).  According to
Eq.~\ref{eq:IF_trans}, the transmitted reflectance peaks strongly when
$\mu < \tau < \mu_o$, and since at ring-plane crossing $\mu$ is very
small, the computed reflectance is very sensitive to optical depths
near zero, so we must be cautious about the portions of the profile
where
the optical depth is very small. Typical noise levels in the Lick
observations of the flux of the occulted star were
$\sigma(F_{obs})/F_o$=0.025, where $F_{obs}$ was the observed flux of
the occulted star and $F_o$ its its unocculted flux. The computed
optical depth is
\begin{equation}
\tau=-\mu \ln \left(\frac{F_{obs}}{F_o}\right),
\end{equation}
where $\mu=\sin B_e=0.429$. \citep{N00}. Thus the 2.5\% noise level results in
an uncertainty in $\tau$ of 0.01. Whenever the optical depth value
is less than this, it is set to zero to eliminate spurious peaks in
the computed reflectance profiles. 

In the Voyager observations, which span phase
angles of 6$^\circ$--155$^\circ$, the rings obey a power-law phase
function which is suitable for macroscopic particles:
\begin{equation}
P(\alpha)= c_n  (\pi - \alpha)^n,
\label{eq:phase_function}
\end{equation}
where $c_n$ is a normalization constant.  The phase function of the A
ring is similar to the phase function of Callisto, with $n=3.301$ and
$c_n=0.130$. \citep{D93} 

This is the phase function we adopt for the
main rings, despite the low phase angle of the HST observations.

We interpolate the ring opening angles to the Sun and Earth from
ephemerides from the Planetary Data System's Rings Node ({\tt
  http://pds-rings.seti.org}) using the nominal time for each HST visit
in Table~\ref{table:ring_opening}.

\subsection{The F-ring model}

The F~ring of our model is constructed as an inclined ``ribbon'' of
cells that are smaller than the main ring pixels.  Each cell's radial
optical depth is determined using a Gaussian profile with height given
by Eq.~\ref{eq:tau_h}.  The F~ring is divided into a front half and back half.   We create positive-valued 2-D images of the sunlight reflected from the front (FF) and back (FB), and negative-valued images of light absorbed or scattered by the F~ring from the back of the F~ring (FBF) and the main rings (FBM).

\subsubsection{Constructing the F-ring model}
To determine the geometry of the F~ring at the time of these
observations, we must take into account the regression of the F~ring's
ascending node.  Its longitude is measured from the ascending node of
Saturn's equatorial plane on the Earth's equatorial plane of
J2000.0. The orientation of Saturn's equatorial plane is derived from
the position of its pole, which was not available from the Rings Node
ephemeris, so we compute the right ascension and declination of
Saturn's pole ($\alpha_p,\delta_p$) using the pole and precession rate
of Saturn derived by \citet{F93}.  Combined with Saturn's right
ascension and declination, this allows us to find the position of the
ascending node (on the Earth's equatorial plane) of Saturn's
equatorial plane relative to the line of sight to the observer, to which we can refer the ascending node of the F~ring.

The F~ring model is created as a collection of two-dimensional arrays.
There are $m_f=6000$ cells in the azimuthal direction and, coincidentally, $n_f=6000$ cells in the
vertical direction.  Each cell is assigned an azimuthal coordinate
$\theta_f$, relative to the line of sight to the observer projected
into the F-ring plane, and a vertical coordinate $h$. The distance from Saturn is
taken as a constant, equal to the semimajor axis, $a_f$. Each cell is also assigned a radial optical
depth $\tau_r$ based on its $h$ value.

Once the arrays have been constructed, the cells can be positioned in
space by transforming their coordinates.   Using the F~ring's
inclination and the orientation of its ascending node, the array
containing cylindrical F-ring coordinates, $(r_f,\theta_f,h)$ for each
cell, can be transformed by matrix rotation to Saturn equatorial
coordinates and then to sky-plane coordinates. The origin of the
coordinate systems is the center of Saturn.  The $\hat x$ axis of
Saturn equatorial coordinates is the projection of a vector toward
Earth into Saturn's equatorial plane, the $\hat y$ axis points
westward in Saturn's equatorial plane and the plane of the sky, and
$\hat z$ is aligned with Saturn's north pole.  For the sky-plane
coordinates, $\hat u=\hat y$, and $\hat v$ is aligned with the
projection of $\hat z$ into the sky plane.

We split the F~ring into the back and front halves by making
separate lists of cells with $x<0$ and $x>0$ respectively.

The product of the albedo and phase function of the F~ring are initially set to
$P(\alpha) \varpi_o =1$. The F-ring brightness and will be scaled later to
match the observed dark-side brightness of the rings.

The main rings are treated as thin and flat so
that $\mu=\sin|B_e|$ for all points on the ring, but the F~ring is
modeled as a vertical surface, so that
\begin{equation}
\mu\sim\mu_o=\cos(\theta_f)
\label{eq:theta_f_ring}
\end{equation}
where $\theta_f$ is measured in Saturn equatorial coordinates from the
sub-Earth point on the F ring.  Thus the I/F of each F-ring cell due to reflected sunlight is computed using
Eq.~\ref{eq:IF_ref}:
\begin{equation}
{\rm I/F}= \frac{1}{8} P(\alpha) \varpi_o \left(1-e^{-2\tau_r(h)/\mu(\theta)}\right).
\end{equation}
\label{sec:f_ring_if}

Multiplying by the projected area of the
cell, $dh=\frac{h_f}{n_f}\frac{2 \pi r_f}{m_f}|\mu|$, yields
the AIF of each cell.  (We must take the absolute value of the cosine
because the $\theta$ values for the back of the F~ring are between
$\pi/2$ and $3\pi/2$.)

To create an image of the F~ring's brightness, we first create a
standard 500$\times$4000-pixel  image array with a value of zero for
all pixels. The AIF of the reflected sunlight of each cell in the
front of the F~ring is then added to the appropriate pixel, creating the FF image.  

The same process is used to create an image of the sunlight reflected
from the back half of the F~ring, BF, except that the I/F of the pixels where the back of the F~ring lies behind the main rings is set to zero.  (The main ring opening angle is very small, and even in the C~ring and Cassini Division, $\tau/\mu\sim0.02/0.008=2.5$.  Therefore light reflected from the back half of the F~ring and transmitted through the main rings is negligible.)

\label{sec:f_ring_blocking} 
We also create two image layers to represent the light blocked by the
front of the F~ring: the blocked light from the back of the F ring
(FBF) and the blocked light from the main rings (FBM). For each of
these,  we create an image array with all the pixel values set to
zero.  Because F-ring cells are smaller than u-v pixels, we simply use
the $(u,v)$ coordinate of the center of each F-ring cell to determine
what pixel of the blocked image is behind the cell.  Then the I/F of
that pixel is multiplied by the factor $1-e^{-\tau_r/{\mu}}$ to find
the I/F$_{b}$ that is blocked by the cell.  This is multiplied by the
projected area of the F-ring cell to find the AIF$_b$ absorbed or
scattered by the cell.  The AIF$_b$ blocked by each cell in the front
of the F~ring is then {\em subtracted} from the pixel upon which the
center of the cell falls.  This process is performed on the BF layer
and the MR layer, creating the negative-valued FBF and FBM layers, respectively.

\subsubsection{Albedo scaling}
\label{sec:albedo_scaling}

To begin a complete model run, each of the image layers is computed
for the four
dark-side observation times (14:00--20:00 UT).  During these times,
the brightness contributed by the main ring is much less than the
brightness contributed by the F~ring, so the profile of the total ring
brightness is approximately
\begin{equation}
{\rm VIF}_{tot}(r)\approx{\rm VIF}_{BF}(r)+{\rm VIF}_{FBF}(r)+{\rm VIF}_{FF}(r),
\end{equation}
and also
\begin{equation}
\langle{\rm VIF}\rangle_{tot}\approx\langle{\rm VIF}\rangle_{BF}+\langle{\rm VIF}\rangle_{FBF}+\langle{\rm VIF}\rangle_{FF}.
\end{equation}

Initially, we set  $P(\alpha)\varpi_o=1.$ This provisional model
brightness at each time is then represented by VIF$^\prime$.  The same factor
$P(\alpha)\varpi_o$ is present in the single-scattering equations used to compute the I/F of sunlight reflected by the
front and the back of the F~ring. When the FBF layer is computed, the reflected  light from the back of the F
ring is simply reduced by the factor $e^{-\tau_r/\mu}$, so the FBF layer depends linearly on $P(\alpha)\varpi_o$ as well. The model values therefore can be rescaled
by a factor $p_f$ representing the factor $P(\alpha)\varpi_o$, so that
\begin{equation}
p_f=\langle{\rm VIF}\rangle_{HST}/ \langle{\rm VIF}^\prime\rangle_{tot},
\end{equation}
where $\langle$VIF$\rangle_{HST}$ is the radially-averaged, vertically-integrated I/F measured for this time from the HST observations.

We use the $\langle$VIF$\rangle_{HST}$ values from
Table~\ref{table:HST_VIF_WF3} to compute $p_f$ for each ansa at each
time before the RPX, then average them all together. Each VIF$^\prime_{BF}(r)$, VIF$^\prime_{FBF}(r)$, and VIF$^\prime_{FF}(r)$ is
multiplied by this average $p_f$ to obtain the final model profiles
VIF$_{BF}(r)$, VIF$_{FBF}(r)$, and VIF$_{FF}(r)$.  

The brightness of the lit side of the rings includes significant
contributions from both the F~ring and the main rings.  We have
determined an empirical value of $p_f=P(\alpha)\varpi_o$, which can now be
used in computations of the I/F of the F~ring on the lit side,
but for the main rings, the albedo remains uncertain.  We use the
nominal values of $P(\alpha)\varpi_o$ described in Sec.~\ref{sec:main_ring_model} to compute
VIF$^\prime(r)$ for the main rings, but these must also be
corrected by some factor $p_m$.  The total model brightness on the lit
side of the rings can be expressed as:
\begin{equation}
{\rm VIF}_{tot}(r)={\rm VIF}_{BF}(r)+ p_m {\rm VIF}^\prime_{MR}(r)+ {\rm VIF}_{FBF}(r)
+ p_m {\rm VIF}^\prime_{FBM}(r) + {\rm VIF}_{FF}(r). 
\label{eq:vif_tot}
\end{equation}
On the lit side again, we want the average model brightness to match the HST
brightness, so we set
$\langle$VIF$\rangle_{tot}=\langle$VIF$\rangle_{HST}$ for each ansa at
each time, and solve Eq.~\ref{eq:vif_tot} for:
\begin{equation}
p_m = \frac{\langle{\rm VIF}\rangle_{HST}-\left(\langle{\rm VIF}\rangle_{BF}+ 
\langle{\rm VIF}\rangle_{FBF}+ \langle{\rm VIF}\rangle_{FF}\right)} 
{\langle{\rm VIF}^\prime\rangle_{MR}+ \langle{\rm VIF}^\prime\rangle_{FBM}}.
\end{equation}
This factor, averaged over all the lit-side data points, is then
applied to all computed profiles VIF$^\prime_{MR}(r)$ and VIF$^\prime_{FBM}(r)$,
including those for the dark-side.

We can then finally find VIF$_{tot}(r)$ using Eq.~\ref{eq:vif_tot}.  This profile, and its average over the range $r=80,000$--120,000~km, represent the output of the model, which we then compare to the HST data.

\label{sec:saturnshine} 

\begin{figure}
\begin{center}
\includegraphics[width=5in]{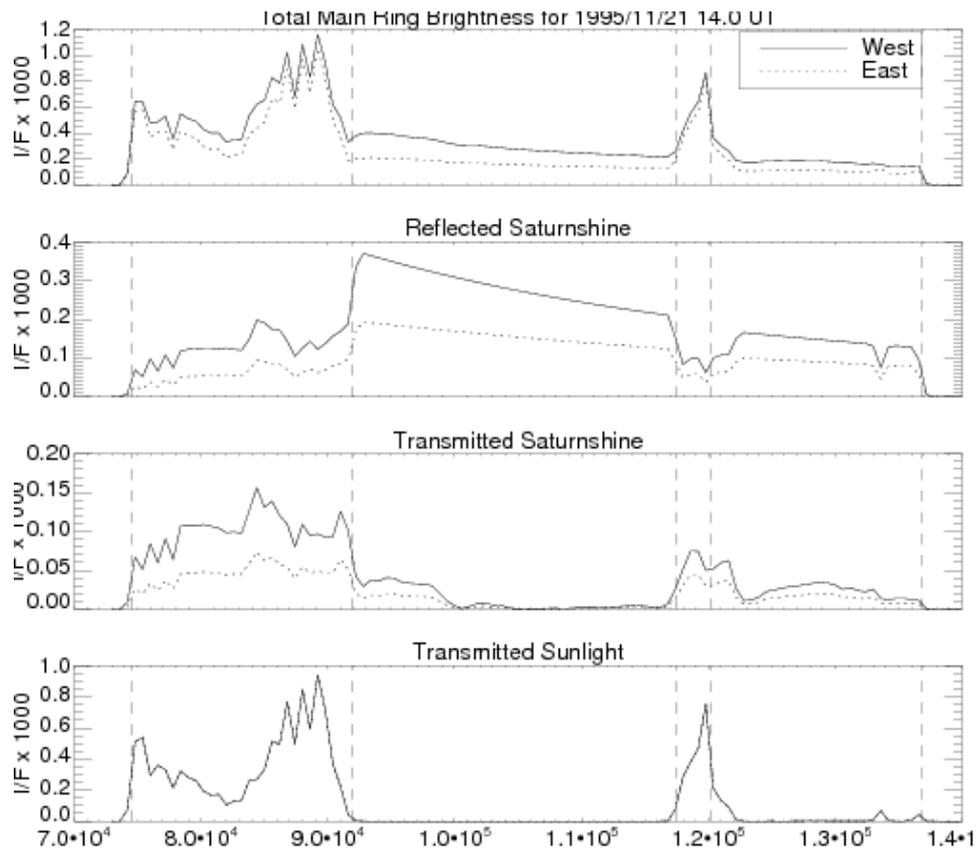}
\end{center}
\caption{A model of the brightness of the east and west ansa as a
  function of distance from the center of Saturn, including
  saturnshine from the northern hemisphere of the planet reflected
  from the rings, and saturnshine from the southern side of the planet
  transmitted through the rings, and sunlight transmitted through the
  rings. The F ring is not included in this model; the asymmetry
  between the east and west ansae comes from the fact that at the time
  of the observation, the phase angle was large and the Sun
  illuminated the western face of Saturn to a greater degree. This
  reproduces the main-ring profiles shown in Fig 4 of \citet{N96}. }
\label{plot:nov_ss}
\end{figure}

The results of the model were verified for simple test cases,
e.g. matching the HST ring brightness profile of \citet{N96} from the
solar ring-plane crossing of 21 November 1995, when, due to the larger
Earth ring-opening angle, the F ring does not obscure the main rings.
The model brightness profile is shown in
Fig.~\ref{plot:nov_ss}, for comparison with Fig. 4 of \citet{N96}. 
At this time, saturnshine was an important source of illumination of
the rings, but
our model shows that the saturnshine was not a significant component
of the ring brightness for the August RPX.  The saturnshine
computation is explained in \citet{S07}.

\bibliographystyle{elsart-harv}
\bibliography{f_ring_inc.bib}

\end{document}